\documentclass[letterpaper,twocolumn,10pt]{article}
\usepackage{usenix2019_v3}
\usepackage{enumitem} 
\usepackage{tikz}
\usepackage{amsmath}
\usepackage{subcaption}
\usepackage[percent]{overpic}

\usepackage{filecontents}

\usepackage[multiple]{footmisc}

\usepackage{multirow}
\usepackage{diagbox}
\usepackage{makecell} %

\usepackage{subcaption}

\newcommand{\para}[1]{\noindent \textbf{#1 }}

\begin{document}

\date{}

\title{\Large \bf Rethinking Network Topologies for Cost-Effective Mixture-of-Experts LLM Serving
}

\author{
{\rm Junsun Choi}\footnotemark[1] \\
UC Berkeley
\and
{\rm Sam Son}\footnotemark[1] \\
UC Berkeley
\and
{\rm Sunjin Choi}\\
UC Berkeley
\and
{\rm Hansung Kim}\\
UC Berkeley
\and
{\rm Yakun Sophia Shao}\\
UC Berkeley
\and
{\rm Scott Shenker}\\
UC Berkeley \& ICSI
\and
{\rm Sylvia Ratnasamy}\\
UC Berkeley
\and
{\rm Borivoje Nikolic}\\
UC Berkeley
} %

\maketitle
\def\thefootnote{*}\footnotetext{Co-first authors}\def\thefootnote{\arabic{footnote}}
\begin{abstract}
Mixture-of-experts (MoE) architectures have turned LLM serving into a cluster-scale workload in which communication consumes a considerable portion of LLM serving runtime. This has prompted industry to invest heavily in expensive high-bandwidth scale-up networks.
We question whether such costly infrastructure is strictly necessary.
We present the first systematic cross-layer analysis of network cost-effectiveness for MoE LLM serving, comparing four representative XPU (e.g., GPU/TPU) topologies (scale-up, scale-out, 3D torus, and 3D full-mesh).
We find that lower-cost switchless topologies are more cost-effective than the scale-up topology across all serving scenarios explored, improving cost-effectiveness by 20.6–56.2\%. In particular, the 3D full-mesh topology is Pareto-optimal in terms of the performance-cost tradeoff.
We also find that current scale-up link bandwidths are over-provisioned: reducing the link bandwidth improves throughput per cost by up to 27\%.
A forward-looking analysis of upcoming GPU generations indicates that the cost-performance advantage of switchless networks will likely persist.
\end{abstract}

\section{Introduction} \label{sec: intro}
\begin{figure}[t!]
    \centering
    \includegraphics[width=0.95\linewidth]{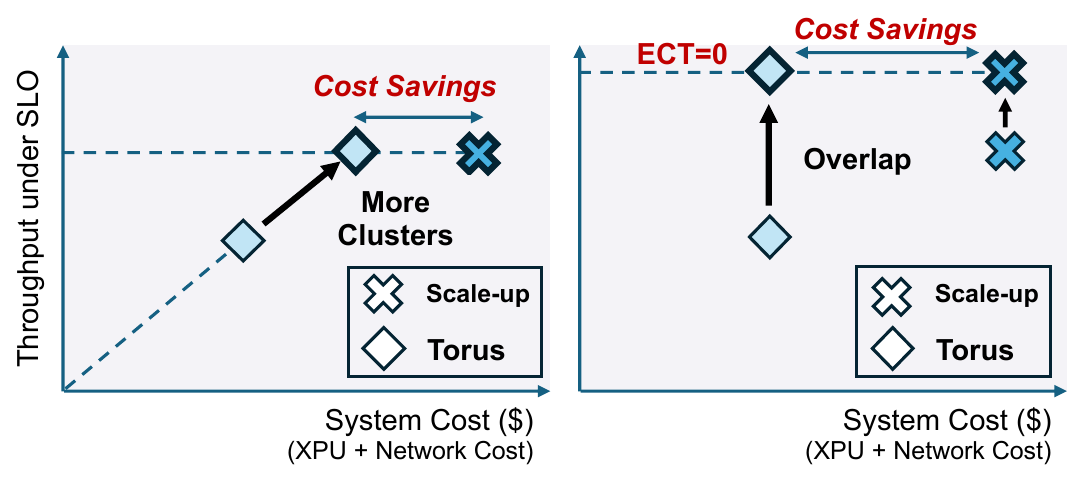}
    \caption{A lower-performance but low-cost network can be a better choice for an MoE serving cluster. Left: A torus cluster may provide lower throughput (smaller Y value) than a scale-up cluster, but its lower cost (smaller X value) allows deploying more clusters, reducing cost at iso-throughput. Right: With computation-communication overlap, the lower-cost topology is the better choice if the exposed communication time (ECT) is zero in both clusters.}
    \vspace{-0.1in}
    \label{fig:marketing}
\end{figure}

As the adoption of LLMs grows, from code generation \cite{claude-code} to scientific discovery \cite{alpha-evolve}, LLM serving is becoming increasingly important. OpenAI serves more than 250 million tokens per second \cite{openai-tokens-per-min}. Hyperscalers are investing hundreds of billions of dollars in GPU clusters to meet this demand \cite{ai-infra-cost}.

With the rise of mixture-of-experts (MoE) architectures in LLMs, LLM serving has recently evolved from a single-node workload into a cluster-scale workload. The MoE architecture keeps per-token computation relatively bounded, enabling model capacity to scale to even trillions of parameters. This rapid model size growth outpaces single-device memory capacity growth, exacerbating the ``memory wall" problem \cite{memory-wall}, forcing the number of XPUs (chips to run AI models on, such as GPUs or TPUs) required for serving to reach cluster-scale.

As serving becomes parallelized, communication accounts for a considerable portion of the end-to-end runtime of LLM serving \cite{mad-max, llm-bottleneck}. For example, communication consumes up to 59\% of runtime in a 96-GPU MoE serving setup~\cite{lmsysDeployingDeepSeek}. Industry has reacted by building ever-larger "scale-up" networks~\cite{nvl72,nvl576}, enlarging the high-bandwidth domain to minimize the communication time,
but also inflating cluster costs with numerous expensive high-capacity switches.

Is this strategy of increasing network investment necessary?
Can a lower-cost network still be sufficiently performant for MoE LLM serving?
We present such a scenario in Figure~\ref{fig:marketing}.
An alternative is to use a lower-cost switchless topology like a torus or lower the link bandwidth, which could be slower but still be preferable,
since the saved cost allows deploying additional clusters to match the scale-up throughput (Left).
Thus, network designs should be evaluated by their throughput per cost when cluster expansion is feasible.
Also, because LLM serving allows aggressive overlap of communication and computation~\cite{deepseek-dbo,megascale-infer}, exposed communication time (ECT) is the right metric for comparing networks. We need only enough network capacity to achieve ECT$\approx$0: any less hurts performance, and any more wastes investment (Right).

We refer to a network that attains this sweet spot as a “cost-effective” network. In this paper, we focus on estimating the cost-effectiveness of different network designs. 

Determining the sweet spot is challenging because it requires a \textit{holistic} view of performance. That is, one must look beyond network performance in isolation and instead understand both compute and communication times \emph{as well as} how they overlap. For this, we must understand how key system parameters such as memory size, context length, and service-level objective (SLO) requirements jointly impact performance. 
Furthermore, the cost-performance tradeoff is sensitive to software optimizations like speculative decoding (SD) and dual-batch overlap (DBO) (\S\ref{ssec: sw-opt}) and hence must also be captured by our evaluation.

\begin{figure}[t]
    \centering
    \includegraphics[width=1\linewidth]{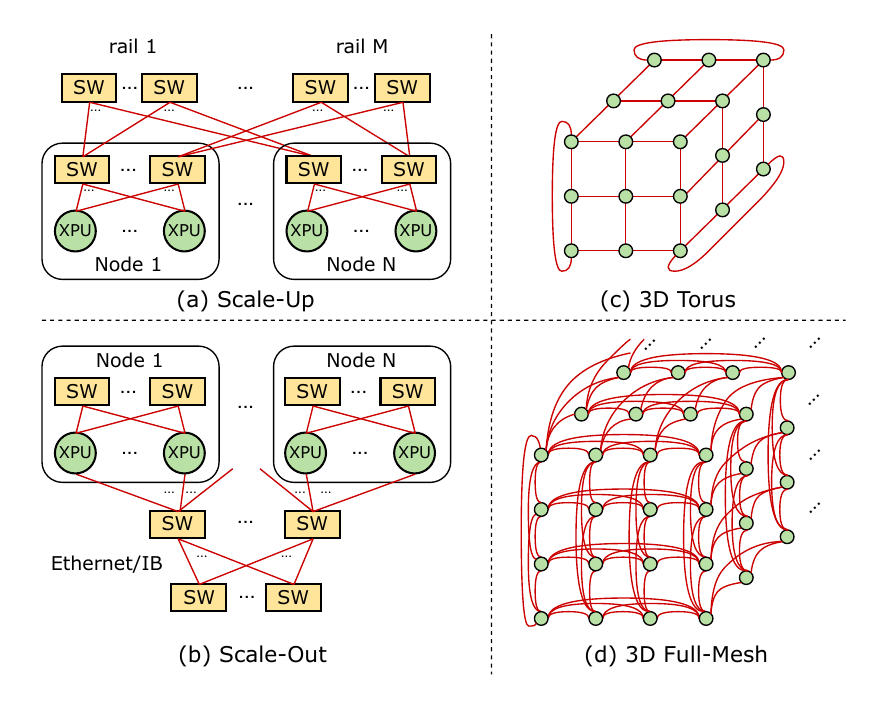}
    \vspace{-0.15in}
    \caption{Network topologies of current XPU clusters}
    \vspace{-0.1in}
    \label{fig:four-topologies-intro}
\end{figure}

We address these challenges for serving via a holistic, cross-layer methodology
that systematically models communication time, compute time, and cluster cost for MoE LLM inference by carefully combining real-world benchmarks with simulation. We apply our methodology to four representative XPU network topologies (scale-up, scale-out, 3D torus, and 3D full-mesh)
\cite{nvl72, tpuv4, ub-mesh}
(Figure~\ref{fig:four-topologies-intro}), analyzing performance-cost tradeoffs in realistic serving scenarios with SD and DBO, and across a range of configurations for link bandwidth and latency.
Our key findings are as follows:

\textit{1) Current systems are not cost-effective:}
We find that in scale-up systems, the current link-bandwidth provisioning lies beyond the sweet spot for cost-effectiveness: reducing link bandwidth alone improves throughput per cost by up to 27\% (\S\ref{ssec:bw-sweep}).

When combined with the topology comparison, our analysis shows lower-cost switchless topologies, such as 3D torus and full-mesh, are more cost-effective than scale-up and scale-out networks across all LLM serving scenarios we explore, improving cost-effectiveness by 20.6–56.2\%. In particular, 3D full-mesh, despite being less widely adopted than other topologies, provides the best overall tradeoff (\S\ref{ssec:result-all},\S\ref{ssec:best-config}).

This trend arises because, under a short context length and a more relaxed SLO (higher TPOT), DBO is effective such that the exposed communication time can be eliminated. SD further expands the regime in which DBO is effective, strengthening the cost-effectiveness of lower-cost switchless networks. Under a long context length or a tight SLO, the small maximum achievable batch size leads to a smaller message size, removing the performance advantage of scale-up over torus and full-mesh.

\textit{2) Switchless topologies will remain more cost-effective than switch-based topologies in the next generation:}
Through a forward-looking analysis that considers projected compute and memory scaling over the next two GPU generations (NVIDIA Blackwell and Rubin), we show that the cost-effectiveness advantage of the switchless networks is likely to persist in the near future for most serving scenarios (\S\ref{ssec:future-projection}). One caveat, however, is that compute performance and memory bandwidth are expected to scale faster than interconnect bandwidth in the Rubin generation (Table~\ref{table:gpu-scaling}). If this trend continues, bandwidth provisioning may be insufficient to sustain the advantages of switchless networks.
Lowering the software and protocol overhead, alongside bandwidth scaling, helps maintain the advantage of switchless topologies in the next generation.

To our knowledge, we are the first to conduct a systematic analysis of the cost-effectiveness of different network designs for MoE LLM serving, taking communication time, compute time, software optimizations, and cluster cost into consideration.
A line of work on network comparison ~\cite{taccl,tacos,themis,swing,halfring,supermesh,rethinking-ml-collective} optimizes the collective communication time (CCT) of LLMs, but not ECT.
Other work has looked at cross-layer XPU network design in the context of training~\cite{alibaba-hpn, ub-mesh, topoopt, crux, autoccl}, but serving presents distinct challenges. First, it uses higher expert parallelism and smaller batch sizes to meet latency targets, leading to different traffic patterns—e.g., more inter-node traffic versus the intra-node dominance in training~\cite{ub-mesh}. Second, serving must satisfy diverse latency SLOs (e.g., TPOT), whereas training primarily optimizes throughput or MFU, requiring scenario-specific topology evaluation. Finally, serving-specific optimizations such as speculative decoding~\cite{spec-decoding,medusa,eagle,big-little} further alter traffic volume and overlap behavior significantly.

In summary, the contributions of our work lie in a \emph{methodology} for evaluating the cost-effectiveness of network designs for serving and our \emph{results} on current state-of-the-art solutions. We believe these contributions will aid system designers in exploring the design space for future cost-efficient MoE LLM serving clusters.

\section{Background} \label{sec: background}

In this section, we explore the unique communication pattern in Mixture-of-Experts serving, and how it has driven the design of network topologies deployed in today's cluster-scale XPU systems (Figure~\ref{fig:four-topologies-intro}). Lastly, we discuss software-level optimizations that can alleviate the high communication demand without additional hardware cost.

\subsection{Communication Pattern in MoE Serving} \label{ssec: background-comm-pattern}

\para{Mixture-of-Experts} LLMs are based on the transformer architecture \cite{attention-is-all-you-need}, where the model consists of a sequence of decoder blocks. Each decoder block contains two main components: self-attention and a feed-forward network (FFN). 
The Mixture-of-Experts (MoE) architecture~\cite{zhou2022mixture} addresses the scaling problem in LLMs by replacing dense FFNs with multiple smaller FFNs (so-called experts) and a router that selects a small subset of experts for each token. This sparse activation increases the model size without proportionally increasing computation. As a result, many state-of-the-art LLMs now use MoE and contain hundreds of experts~\cite{kimi-k2, deepseek-v3}.

\para{Communication Pattern} MoE serving runs on an XPU cluster and involves two key collective communication operations. First, \textit{all-reduce} (AR) arises from tensor parallelism, where each device computes a partial result, and the results are aggregated across devices. Second, and more importantly, MoE introduces expert parallelism (EP), which distributes different experts across devices. Since each token is assigned to only a small subset of experts, tokens must be exchanged with the devices hosting their selected experts, requiring \textit{all-to-all} (A2A) communication. While we consider both collectives in our analysis, we place greater emphasis on A2A because of its larger share of the communication overhead.

\begin{figure}[t]
     \centering
     \includegraphics[width=1\linewidth]{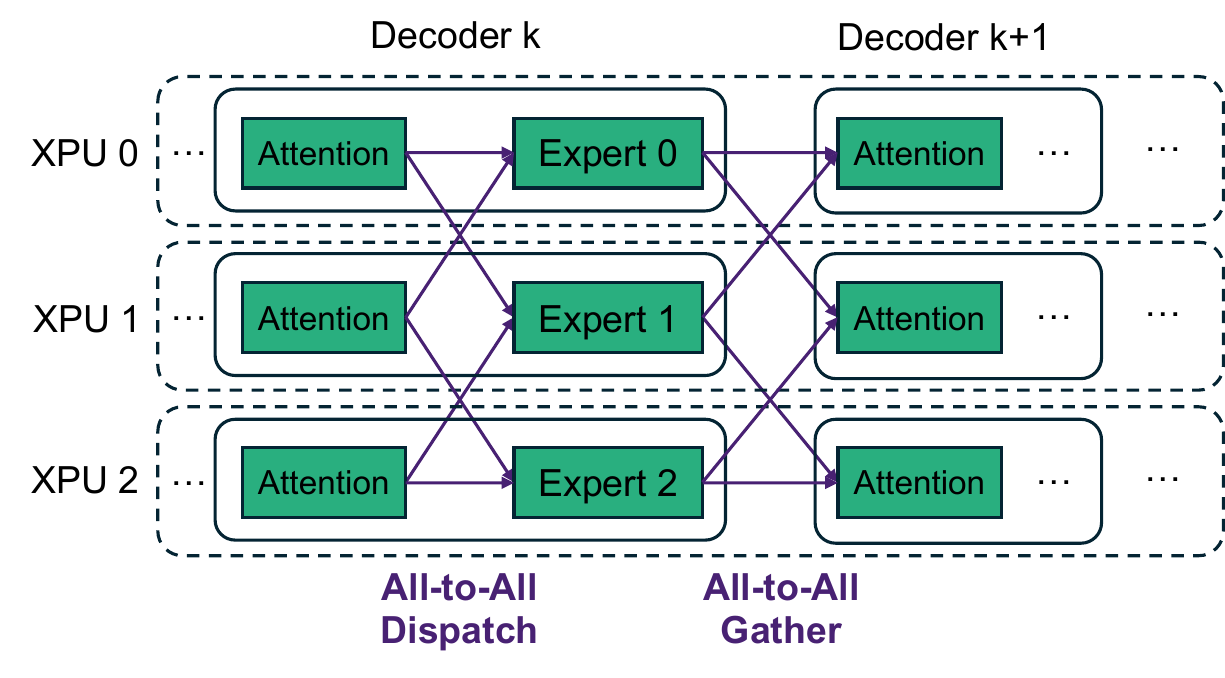}
     \vspace{-0.1in}
     \caption{Communication pattern under expert parallelism. Here, we assume expert parallelism of 3 with one request per XPU, top-2 routing, and omit non-MoE/attention layers. Many MoE models route each token to multiple experts.}
     \vspace{-0.1in}
     \label{fig:comm-pattern}
\end{figure}

Figure~\ref{fig:comm-pattern} illustrates this communication pattern. Each XPU processes its local micro-batch in the self-attention layers. At the boundary between self-attention and the MoE FFN, tokens are shuffled according to the router’s assignments (dispatch). After expert computation, the output tokens are routed back to the XPUs that originated them (gather), performing the second A2A operation.

\subsection{XPU Cluster Network Topologies} \label{ssec: background-topologies}

\para{Clos-based Topologies} 
The conventional choice for XPU clusters today is a Clos-based topology interconnecting GPU servers, which typically hold 8 GPUs each (e.g., NVIDIA DGX), via switched Ethernet or InfiniBand. However, inter-server links offer roughly one-tenth the bandwidth of intra-server interconnects (e.g., 50 ~GB/s vs. 450 ~GB/s for NVLink), limiting A2A bandwidth. Newer systems~\cite{nvl72,amd-helios} mitigate this bottleneck by connecting tens of GPUs in a single high-bandwidth domain. We refer to the former as \textit{scale-out} and the latter as \textit{scale-up}. 
Both use a switch-based, non-blocking fat-tree topology, allowing any XPU to communicate with any other XPU without routing through another XPU or network blocking, which improves collective performance. However, this comes at the cost of many switches and links. The cost grows rapidly in scale-up networks, where per-GPU NVLink bandwidth doubles and the number of GPUs per scale-up domain increases with every GPU generation.

\para{Switchless Topologies} Switchless topologies are an alternative to fat-tree for building XPU clusters. 
We consider two switchless topologies: 3D torus and 3D full-mesh, particularly $4\times4\times4$ configurations with 64 XPUs and $8\times8\times4$ with 256 XPUs. We choose these topologies because they represent real large-scale XPU deployments: 3D torus in Google's TPU clusters and nD full-mesh in Huawei's UB-Mesh-Pod~\cite{tpuv4, ub-mesh}.
As drawn in Fig.~\ref{fig:four-topologies-intro}, a 3D torus connects each node to its six axis-aligned neighbors (three dimensions and two directions) with wrap-around links, while a 3D full-mesh connects each node directly to every other node along each of the three dimensions.

The main benefit of switchless topologies over fat-tree is that they do not require switches, as direct links to the neighbors can be statically configured, saving the cluster network cost.
Both deployments do use switches at scale as OCSes to form wrap-around links across TPU cubes and low-radix switches to sustain inter-rack links in UB-Mesh-Pod, but 
since these merely maintain fixed connections once the topology dimension is set, both can reasonably be considered switchless.

Compared to non-blocking fat-tree, switchless topologies typically achieve lower network cost at the expense of A2A bandwidth. In a torus, distant XPUs communicate over multiple hops, so flows consume multiple links and thus the effective bandwidth per flow is reduced. Full-mesh mitigates this by connecting every XPU in a dimension in one hop, but the fixed off-chip bandwidth is split across more links, reducing per-link bandwidth. Both topologies hence achieve lower effective A2A bandwidth than switch-based alternatives.

\subsection{Software Optimizations for LLM Serving} \label{ssec: sw-opt}

\begin{figure}[t]
    \centering
    \includegraphics[width=0.95\linewidth]
    {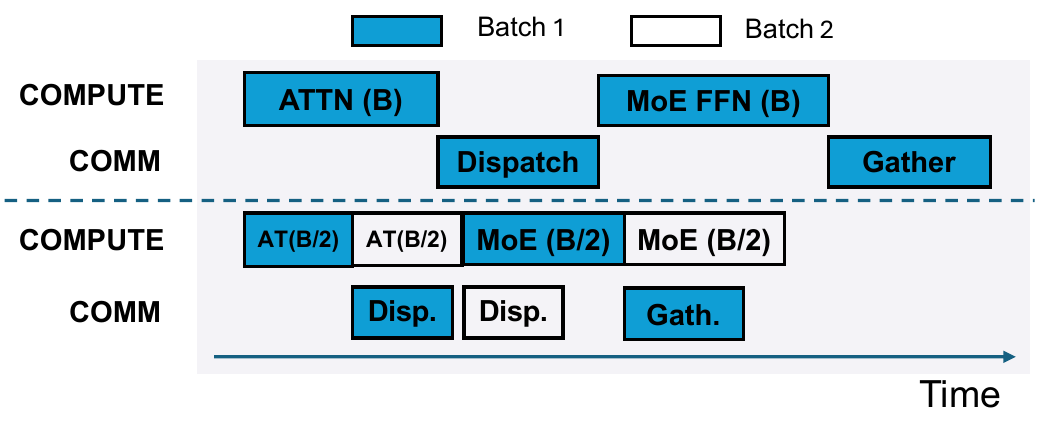}
    \vspace{-0.1in}
    \caption{Top: a standard MoE iteration with batch size B. Bottom: dual-batch overlap.}
    \label{fig:dbo}
\end{figure}

\begin{figure}[t]
    \centering
    \includegraphics[width=0.9\linewidth]{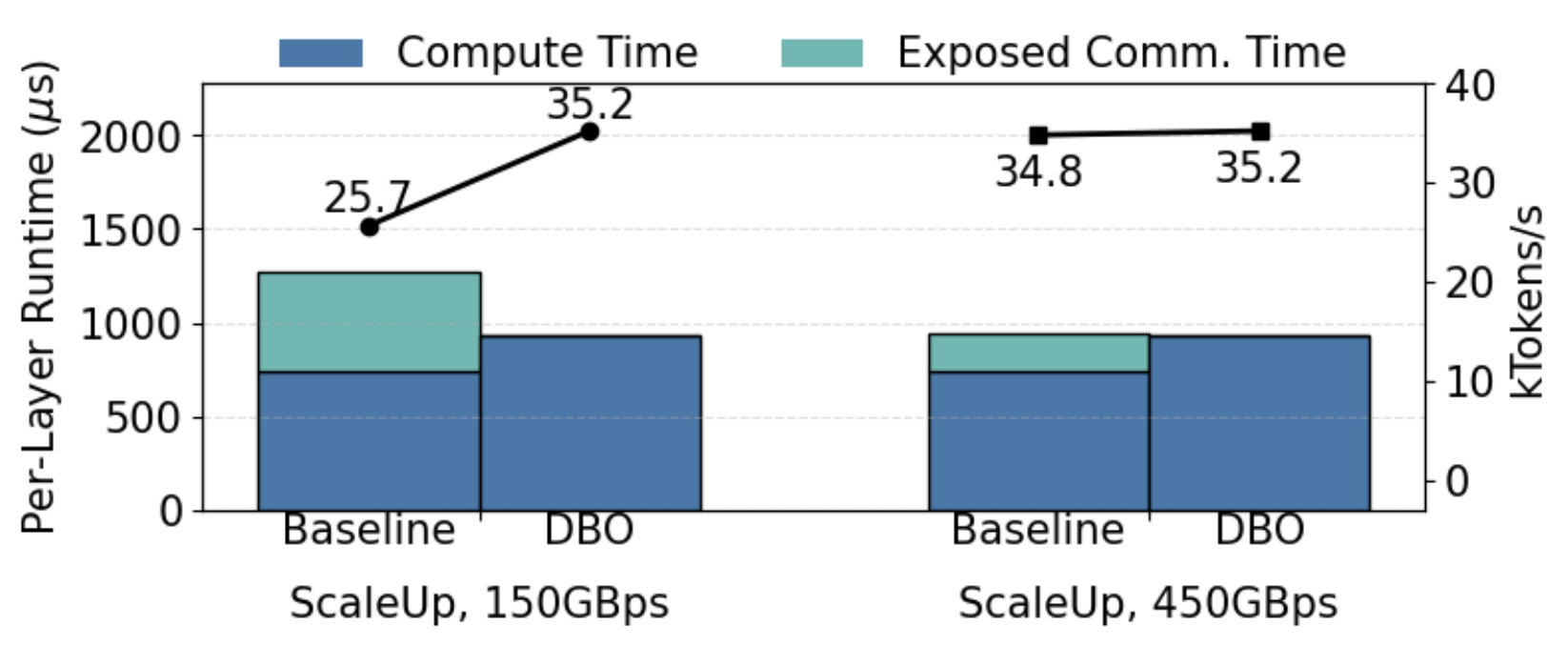}
    \vspace{-0.15in}
    \caption{The per-layer latency and throughput in the two scale-up networks with different link BW (left: lower BW). A lower-cost network can match the performance of expensive networks with compute-communication overlap (DBO).
    Model: DeepSeekV3, EP64, Batch size: 32768, GPU: NVIDIA H100}
    \label{fig:dbo-exp}
\end{figure}

\para{Dual Batch Overlap}
Dual batch overlap (DBO)~\cite{deepseek-dbo,deepseek-v3} is a computation-communication overlap technique for hiding communication time in MoE serving. It splits a batch into two smaller microbatches and overlaps the computation of one microbatch with the communication of the other, as illustrated in Figure~\ref{fig:dbo}. When effective, DBO reduces exposed communication time and end-to-end latency, thereby improving throughput as shown in Figure~\ref{fig:dbo-exp}.


\begin{figure}[h!]
    \centering
    \includegraphics[width=0.9\linewidth]{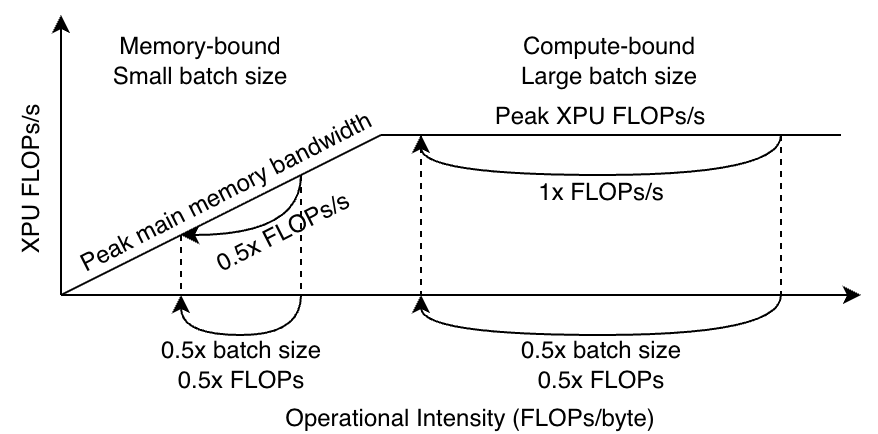}
    \caption{Dual-batch overlap is beneficial only at sufficiently large batch sizes. When batch sizes are small, layers are memory-bandwidth-bound, and splitting the batch into two can lower the throughput.
    }
    \label{fig:dbo-roofline}
\end{figure}

The effectiveness of DBO depends strongly on the operating point: the batch size must be large enough for DBO to be beneficial, and at small batch sizes, it can actually reduce the overall throughput despite driving exposed communication time to nearly zero. Figure~\ref{fig:dbo-roofline} illustrates this effect. A small batch size leads to a lower FLOPs-per-byte ratio for matrix multiplications in LLM layers, such as MoE and Q, K, and V projections, causing them to be bounded by data movement rather than computation.

In this regime, halving the batch halves both FLOPs and achievable FLOPs/s, so each microbatch's compute time barely shrinks. Under DBO, the two microbatches together nearly double the compute time, which can outweigh the communication time that DBO hides and potentially make end-to-end time worse than without DBO.

\para{Speculative Decoding} 
Today’s autoregressive LLMs generate one token at a time, which can be too slow to meet very low TPOT targets such as 10 ms. Speculative decoding (SD)~\cite{spec-decoding,medusa,eagle,big-little} addresses this limitation by trying to generate multiple tokens in a single iteration. It consists of two phases: drafting, where a lightweight predictor proposes several candidate next tokens using either a smaller draft model~\cite{spec-decoding,big-little} or additional decoding heads~\cite{medusa,eagle}, and verification, where the target model evaluates these candidates in parallel and accepts the longest valid prefix. Since one drafting-plus-verification step for N tokens can be faster than N separate autoregressive steps, depending on the operating point, SD can substantially reduce TPOT.

We consider SD because it changes the operating point and the communication pattern of MoE serving. In particular, the verification phase processes multiple candidate tokens at once, effectively increasing batch size and A2A message sizes, which can alter the relative advantage of different network topologies. We detail its effect in \S\ref{sssec:sw-opt}.

\section{Methodology} \label{sec: methodology}
\subsection{Serving Workloads}
In this section, we define the serving workloads, performance target, and evaluation scenarios used throughout our study. 

Among the two phases of LLM serving, prefill and decode, we focus on the decode phase and use the time per output token (TPOT) as the SLO, motivated by two observations. First, decode accounts for most of the end-to-end serving time, especially as demand for long output requests increases~\cite{maliakel2026characterizing,saxena2025utility,zhong2024distserve}. Second, modern serving systems increasingly adopt prefill-decode disaggregation~\cite{zhong2024distserve,patel2024splitwise,nvidiaNVIDIADynamo}, providing a basis for studying the decode phase in isolation.

We restrict our attention to batched inference, which is the common deployment mode for XPU clusters. 
We exclude very long single-request inference to limit our scope to cases where all-to-all communication dominates.

As a representative workload, we use DeepSeek-V3 (671B, 256 experts, top-8 routing)~\cite{deepseek-v3,huggingfaceDeepseekaiDeepSeekV3Hugging}. We focus on a single model because recent large MoE LLMs share broadly similar architectural and communication characteristics.

We evaluate a range of serving scenarios by varying two parameters: TPOT targets and the average context length of requests, defined as the average combined length of inputs and outputs. Although individual requests commonly have different input and output lengths during serving, we focus on their combined average because it affects only the runtime of self-attention layers, which is well approximated by this aggregate measure. We use values representative of public datasets and benchmarks\cite{mlcommons_mlperf_2025, mlcommons_inference_rules_scenarios_2026,kwon2023efficient,zhao2024wildchat,bai2024longwriter}. Specifically, we use context lengths of 512 and 4096 for short- and long-context scenarios, and TPOT requirements of 15 ms, 40 ms, and 100 ms. Representative workloads for each TPOT target correspond to a reasoning model, a chatbot, and synthetic data generation or RLHF training, respectively.

\begin{table}[t]
\centering
\begin{tabular}{lcc}
\hline
 & Intra-node & Inter-node \\
\hline
$\alpha_0$ & 5.874\,\textmu s & 26.508\,\textmu s \\
$\alpha_{r}$ & 0.809\,\textmu s & 1.358\,\textmu s \\
$\alpha_{d}$ & 0.323\,\textmu s & 0.340\,\textmu s \\
Link utilization & 71.7\% & 84.3\% \\
\hline
\end{tabular}
\caption{Alpha-beta model parameters}
\label{table: a-b-values}
\end{table}

\begin{table}
\small
\begin{tabular}{|p{1.4cm}|p{3.2cm}|p{2.7cm}|}
\hline
\diagbox[width=1.8cm]{\hspace{-0.2cm}Topology}{Op} & \makecell[c]{AllReduce} & \makecell[c]{All-to-all} \\ \hline
\makecell[c]{Scale-up\\Scale-out} & \makecell[c]{Recursive doubling, P2P,\\ Rabenseifner, Ring~\cite{gropp}} & \makecell[c]{Bruck, Spread-out,\\P2P~\cite{gropp, bruck, spread-out}} \\ \hline
\makecell[c]{Torus} & Ring, Swing, P2P~\cite{swing,aws-collectives-intra-node}& HalfRing, P2P~\cite{halfring,aws-collectives-inter-node} \\ \hline
\makecell[c]{Full-mesh} & \makecell[c]{Ring, P2P} & \makecell[c]{\hspace{-0.1cm}One-shot, DoR} \\ \hline
\end{tabular}
\caption{List of allreduce and all-to-all algorithms considered for different topologies in communication time modeling}
\label{table: algorithm-list}
\end{table}

\subsection{Performance Modeling} 
\label{ssec: perf-modeling}
Because our study includes hypothetical network topologies and link bandwidths that are not publicly available, we cannot solely rely on direct performance measurement. We therefore use performance modeling to estimate MoE LLM compute and communication time across various clusters, following prior work~\cite{hockney,gropp, bruck, spread-out, swing, halfring,hnp,topoopt}.

\subsubsection{Overview}
\para{Setup} We base our performance model on compute, memory, and link bandwidth specifications on NVIDIA Hopper GPUs to represent the current-generation XPUs. Newer Blackwell GPUs are available but remain scarce and expensive, so Hopper is still the workhorse today. Baseline unidirectional bandwidths per XPU are 450 ~GB/s (scale-up) and 50 ~GB/s (scale-out). In torus and full-mesh topologies, each XPU's scale-up bandwidth is split evenly across its links, e.g., a 3D torus gives 450 / 6 = 75 ~GB/s per link.

\para{Metric} To compare the cost-effectiveness of clusters for batched inference, we use two primary performance metrics: the maximum throughput achievable under a TPOT constraint, corresponding throughput normalized by cost. We use maximum capacity as a metric because cluster builders typically provision clusters for expected \textit{peak load}, often with slight over-provisioning, to avoid overload and maintain SLO compliance. Under this provisioning model, improving maximum cluster capacity per cost translates directly into cluster cost savings.

To model this metric, we sweep all parallelism configurations and batch sizes that satisfy the GPU memory capacity constraint. For each configuration, we compute throughput as batch size divided by TPOT, and then select the configuration with the highest throughput under the given TPOT constraint. TPOT is modeled by aggregating the modeled runtime of communication (\S\ref{sssec: comm-time-modeling}) and compute layers (\S\ref{sssec: comp-time-modeling}). This closely matches real serving systems, as we will show in \S\ref{ssec: validation}, since CUDA graphs eliminate most kernel-launch overhead during the decode phase.

\subsubsection{Communication Time} \label{sssec: comm-time-modeling}

We model AR and A2A communication time using the alpha-beta (Hockney) model~\cite{hockney,taccl,efficient-a2a-direct-connect,rethinking-ml-collective,gropp,swing}, where sending a message of size $m$ takes $T = \alpha + m\beta$, with $\alpha$ as the per-message latency and $\beta$ the reciprocal link bandwidth.
Although expert A2A traffic can be irregular, we assume regular A2A since load-balancing makes it approximately regular at large enough batch sizes~\cite{deepseek-v3,deepseek-eplb,lmsysDeployingDeepSeek}.
To ground the model in real hardware, we fit the alpha-beta model to NCCL AR and A2A runtimes collected on NVIDIA DGX H100 nodes, sweeping message sizes from 128 B to 16 GiB and GPU counts from 4 to 32 in powers of 2.
The measurements motivated three adaptations to fit the model better. First, we add a one-time latency term $\alpha_0$ on top of $\alpha$. Second, we decompose $\alpha$ into a per-communication-round latency $\alpha_r$ and a per-destination serialization cost $\alpha_d$. $\alpha_r$ is added to capture why A2A runtime grows with GPU count. Third, we define $\beta = \text{(link utilization)} \times 1/\text{(peak link bandwidth)}$ and fit link utilization to generalize across bandwidths.

The fitted values are listed in Table \ref{table: a-b-values}. The model shows mean relative errors of 10.82 percent and 7.97 percent compared to the intra-node and inter-node measurements, respectively, where the relative error is defined as (actual runtime - model estimate)/(actual runtime). Our fitted $\alpha$ and link utilization are similar to the measurements in previous literature~\cite{taccl}. Our model relies on GPU measurements, but our methodology can be generalized to other types of XPU. 

Using the fitted parameters, we extend our model across topologies, algorithms, and XPU counts.
Table~\ref{table: algorithm-list} lists the algorithms considered. For fat-tree and torus, we picked widely used or state-of-the-art designs~\cite{gropp,bruck,spread-out,swing,halfring}. P2P (point-to-point, or send-recv) refers to direct pairwise exchange. In P2P A2A, each XPU sends a unique message to every other XPU, as in NCCL. For full-mesh, the one-shot and DoR (dimension-order routing~\cite{tpuv4}) A2A algorithms adapt the torus P2P and HalfRing~\cite{halfring} to full-mesh, respectively. We assume hardware-assisted cut-through routing, as evidenced by its use in AWS Trainium systems~\cite{aws-collectives-intra-node,aws-collectives-inter-node}.

\begin{table}[t]
\centering
\small
\setlength{\tabcolsep}{4pt}
\begin{tabular}{l|c|c}
\hline
\textbf{Algorithm} & $N=64$ (4$\times$4$\times$4) & $N=256$ (8$\times$8$\times$4) \\
\hline
ScaleUp-P2P 
& $1\alpha_{r} + 63\alpha_{d} + \frac{63}{64}m\beta$ 
& $1\alpha_{r} + 255\alpha_{d} + \frac{255}{256}m\beta$ \\

ScaleUp-Bruck 
& $6\alpha_{r} + 6\alpha_{d} + 3m\beta$ 
& $8\alpha_{r} + 8\alpha_{d} + 4m\beta$ \\

FullMesh-DoR 
& $3\alpha_{r} + 27\alpha_{d} + \frac{9}{4}m\beta$ 
& $3\alpha_{r} + 51\alpha_{d} + \frac{17}{4}m\beta$ \\

Torus-HalfRing 
& $6\alpha_{r} + 36\alpha_{d} + 3m\beta$ 
& $12\alpha_{r} + 72\alpha_{d} + 6m\beta$ \\
\hline
\end{tabular}
\caption{$\alpha_{r},\alpha_{d},\beta,$ coefficients for selected A2A algorithms under different cluster sizes. Lower coefficient values mean shorter collective completion time. $N$ is the number of XPUs.}
\label{table:a2a-alg-coefficients}
\end{table}

\begin{figure}[t]
    \centering
    \includegraphics[width=\linewidth]{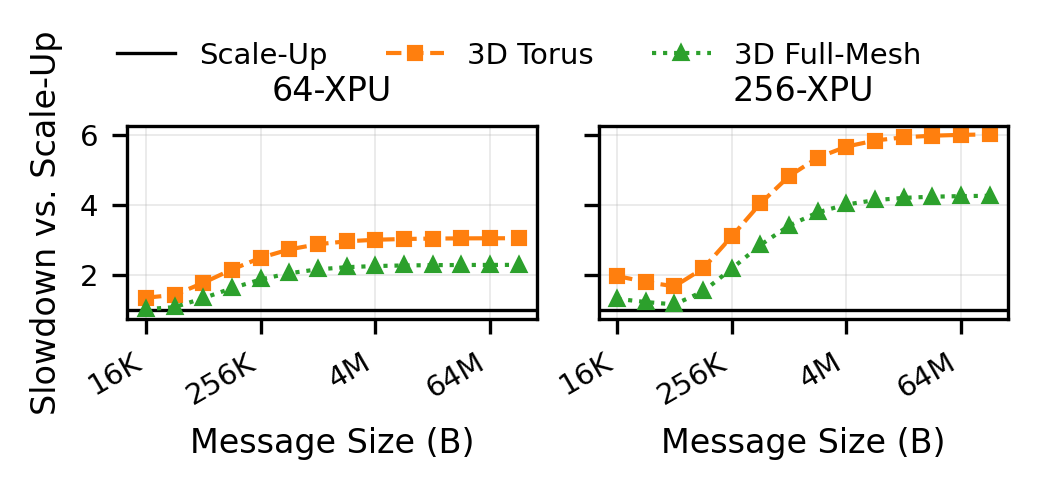}
    \caption{A2A communication time of each topology}
    \label{fig:a2a-alg-comparison}
\end{figure}

For each cluster configuration (topology and XPU count), we select the algorithm minimizing AR and A2A time as our communication estimate. For the A2A algorithms in Table~\ref{table:a2a-alg-coefficients}, Figure~\ref{fig:a2a-alg-comparison} shows the fitted alpha-beta model. As discussed in Section~\ref{ssec: background-comm-pattern}, scale-up achieves the best A2A performance in terms of both $\alpha$ and $\beta$ terms, and full-mesh beats torus on both terms thanks to its higher connectivity. 

To validate that our model generalizes to algorithms beyond those measured, we implemented ring and recursive doubling AR as custom NCCL kernels. Ring AR serves as a control since NCCL provides it natively. Our implementation outperforms NCCL's, so substituting it does not compromise representativeness. Fitting each algorithm's alpha-beta model to runtimes measured on an AWS p4.24xlarge instance (8$\times$ NVIDIA A100) yields only 4.79\% and 3.32\% mean relative errors for ring and recursive doubling, respectively, confirming that the model applies broadly across algorithms.

\subsubsection{Compute Time} \label{sssec: comp-time-modeling}
High-fidelity serving performance modeling has been extensively studied~\cite{agrawal2024vidur,topoopt,cho2024llmservingsim,lin2024apex}, so we leverage existing methods to model MoE serving. Among these approaches, we adopt a profile-based method because it captures runtime more accurately than an analytical roofline-based model~\cite{williams2009roofline}. In particular, we use a single-GPU measurement methodology similar to Vidur~\cite{agrawal2024vidur}. Under tensor parallelism and expert parallelism, each GPU within a parallelism domain executes a shard with the same tensor shape. Leveraging this property, we derive the tensor shapes for each layer under different parallelism configurations, profile kernel performance on a single GPU for each setup, and then use these measurements to estimate the compute time of each layer. Since Vidur is designed only for dense LLMs, we further extend this approach to support expert parallelism and MoE architectures.

\subsubsection{Validation of End-to-End Runtime Estimation} \label{ssec: validation}
\begin{figure}[t]
    \centering
    \includegraphics[width=1\linewidth]{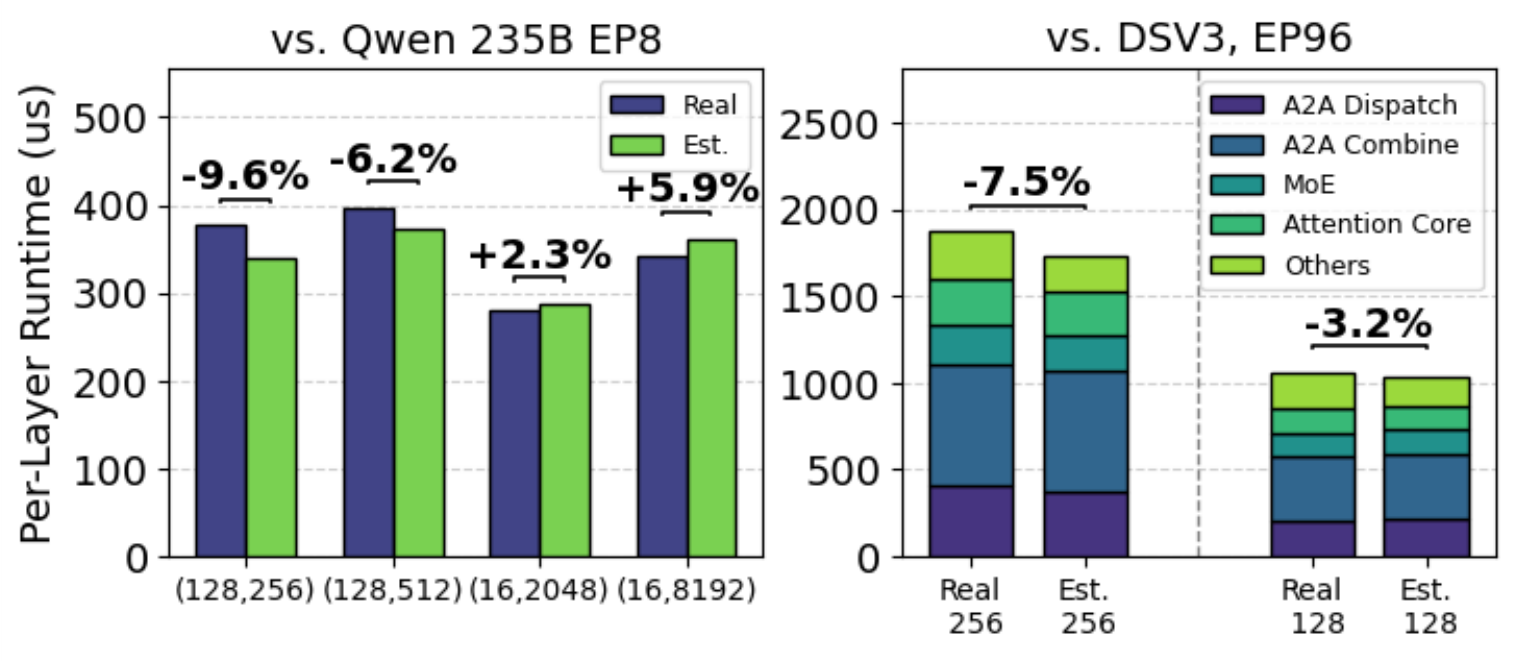}
    \vspace{-0.2in}
    \caption{Validation of runtime estimation. Left: Comparison against measurement on an HGX (8$\times$H100). X-axis: (batch size, context length)
    GPUs.
    Right: Comparison against a public layer-level runtime trace from SGLang on 16 HGX. X-axis: Batch size. Decode performance is reported.}
    \vspace{-0.1in}
    \label{fig:validation}
\end{figure}

We validate our estimation framework in two settings. First, we compare our predictions against real-world measurements from Clos-based deployments, the only systems we had direct access to. Second, we compare against a public performance report for DeepSeek-V3 on 96 H100 GPUs to validate at a larger scale. As shown in Figure~\ref{fig:validation}, relative errors are below 9.6\% and 7.5\%, respectively, confirming the fidelity of our framework.

\subsection{Software Optimization Modeling}
We model the TPOT when DBO is applied as the sum of the compute time of the two micro-batches and the exposed communication time. For the compute time, we use the result at batch size $B/2$ and multiply it by 2. For the exposed communication time, we retrieve the compute and AR/A2A times at batch size $B/2$ and simulate a simple greedy schedule that respects dependencies within each micro-batch. Each compute or communication layer, is scheduled as soon as its dependencies are resolved and the corresponding compute or communication lane becomes available. The communication time not hidden under this schedule is taken as the exposed communication time.

For speculative decoding, we assume a multi-head variant such as Medusa~\cite{medusa,eagle}. Let $\text{spec}_m$ denote the number of drafted tokens and $\text{spec}_p$ the average pass rate. We model
\[\footnotesize
\mathrm{TPOT} = \frac{t_{\mathrm{draft}} + t_{\mathrm{verify}}}{\text{spec}_m \cdot \text{spec}_p}.
\]
Here, $t_{\mathrm{draft}}$ is modeled as one normal decode iteration. For $t_{\mathrm{verify}}$, we model it as an iteration where the attention layers have a query sequence length of $\text{spec}_m$, while for the other operations we increase the batch size by $\text{spec}_m$. Unless specified otherwise, we use $(\text{spec}_m, \text{spec}_p) = (4, 0.8)$.

\subsection{Cluster TCO Modeling} \label{ssec:cost-modeling}
To calculate monthly TCO, we amortize the CapEx of XPU, switch, and link over a three-year lifetime and add the monthly CapEx and OpEx, following prior literature~\cite{hnp}.

Consistent with prior work~\cite{topoopt, mixnet}, we model CapEx using publicly available catalog prices for GPUs and network equipment. Switch cost is modeled as a linear function of switch capacity (= radix$\times\text{(per-port bandwidth)}$), capturing the intuition that the same switch ASIC can be reconfigured across radix-bandwidth tradeoffs~\cite{Broadcom-SUE}. This model fits observed prices well ($R^2=0.93$). We assume a switch radix of 64 and 16 scale-up and 1 scale-out ports per XPU, with torus and full-mesh treated as switchless. Link cost is modeled as a fixed cost per unit bandwidth by link type, as pricing data shows it is nearly length-independent. Copper and active optical cables (AOC) are used for intra- and inter-rack links respectively, with AOCs priced at 6.7$\times$ that of copper.
We assume 64 XPUs can be placed in one rack.
Our CapEx estimation of the cost of an 8-GPU DGX system falls into the range of the retail price.

For calculating the OpEx, we multiply the thermal design power (TDP) of the equipment by the electricity cost and the power usage effectiveness (PUE) value of AI clusters~\cite{energy-report}.

We acknowledge that the CapEx data based on retail price may overestimate or underestimate the ratio of the network cost in the system. Considering this, we introduce an adjustment factor $c$ such that $\text{(Monthly TCO)} = \text{(Monthly GPU Cost)} + c \times \text{(Monthly Network Cost)}$ and vary $c$. We also acknowledge that OpEx data beyond electricity costs, such as real estate and facility maintenance, is difficult to obtain and may not be fully captured in our model. Throughout this paper, we report costs normalized to a reference unit cost rather than absolute dollar figures.

\section{Analysis}
\begin{table}[t]
\centering
\small
\begin{tabular}{p{2.8cm}p{4.8cm}}
\hline
\textbf{Metric} & \textbf{Key relationship} \\
\hline
Throughput & $= \text{batch size} / \text{TPOT}$ \\ \hline
TPOT & $= t_{\text{comp}} + t_{\text{comm}}$ \\ \hline
$t_{\text{comm}}$ & \makecell[l]{$= \alpha\text{-term} + \text{coeff}\cdot m \beta$; \\ $m$ grows linearly with batch size} \\ \hline
$t_{\text{comp}}$ (small batch) & $\approx$ const (memory-bandwidth-bound) \\
$t_{\text{comp}}$ (large batch) & $\propto$ batch size (compute-bound) \\ \hline
\makecell[l]{Max Batch Size ($B$)} &
\makecell[l]{$t_{\text{comm}}(B) + t_{\text{comp}}(B) \leq \text{TPOT}$ \\ 
[4pt]
$\text{ModelShard} + \text{KV\$}(B) < \text{Memory Cap.}$ \\ 
($\text{KV\$ size} \propto B \cdot \text{context length}$) \\ } \\
\hline
\end{tabular}
\caption{Summary of key relationships governing throughput, latency, and batch size in the decode phase.}
\label{table:key-relationships}
\end{table}

\subsection{First-Order Performance Trend}
Before performing an in-depth network comparison, we introduce a set of rules of thumb that help us understand
batched-inference performance at a high level, and then apply them to different
serving scenarios and software optimizations to explain their effect on
performance and network cost.

\begin{figure}[t]
    \centering
    \includegraphics[width=0.92\linewidth]{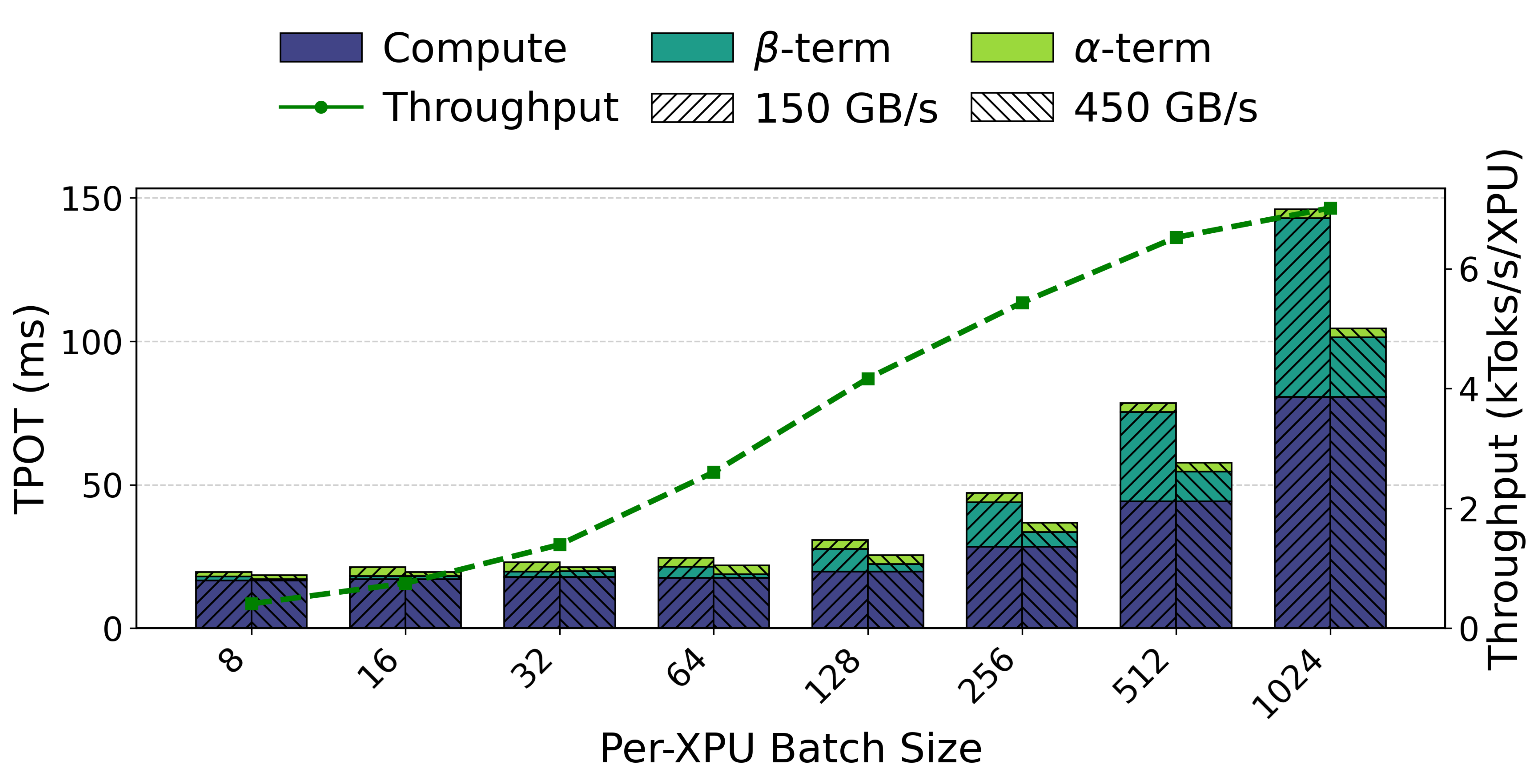}
    \vspace{-0.1in}
    \caption{Latency and throughput as batch size increases in the two scale-up clusters. Context length = 512. Latency (bars) for the clusters with 150~GB/s and 450~GB/s link bandwidths are plotted to show the difference in $\beta$-term. The throughput curve uses 450~GB/s.}
    \label{fig:bs-trend}
\end{figure}

\subsubsection{Batch Size, Latency, and Throughput}
Batch size—the number of user requests processed in the same iteration—is the central quantity governing the tradeoff between latency and throughput in batched inference. Table~\ref{table:key-relationships} summarizes the key relationships:

\begin{itemize}[leftmargin=6pt,itemsep=0pt,parsep=0pt,topsep=0pt,partopsep=0pt]
    \item \textbf{A larger batch size increases TPOT, but only sublinearly at small batch sizes:} $\text{TPOT} = t_{\text{comp}} + t_{\text{comm}} $, and both terms grow with batch size: $t_{\text{comp}}$ from added FLOPs, $t_{\text{comm}}$ from larger messages. However, at small batch sizes, the increase is mild, since the $\beta$-term contributes little and computation is still largely memory-bandwidth-bound. The bars in Figure~\ref{fig:bs-trend} show this trend.
    
    \item \textbf{A larger batch size improves throughput:} Throughput is (batch size) / TPOT, and TPOT grows sublinearly with the batch size. The curve in Figure~\ref{fig:bs-trend} shows this trend.

    \item \textbf{The maximum batch size is limited by TPOT or memory capacity:} The TPOT limit is discussed above. Furthermore, memory capacity limits the KV-cache footprint, which scales with context length and batch size, forcing smaller batches in longer-context scenarios.
\end{itemize}

\subsubsection{Effect of Serving Scenarios}
\label{sssec:effect-serving-scenarios}
\begin{figure}[ht]
    \centering
    \includegraphics[width=1\linewidth]{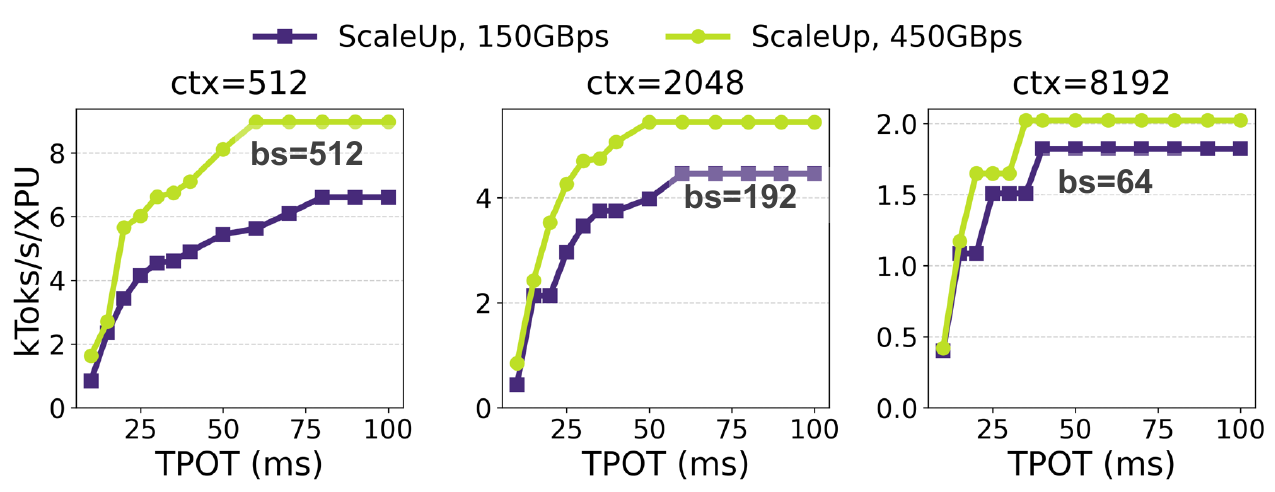}
    \vspace{-0.2in}
    \caption{Effect of TPOT and context length on the two scale-up clusters' token throughput/XPU. Annotations mark the batch size at which GPU memory becomes full.}
    \label{fig:tpot-cl-trend}
\end{figure}
We next show how the two serving scenario parameters, TPOT and context length, affect the performance in different cluster networks. We use two 64 XPU scale-up clusters with 150~GB/s and 450~GB/s link BW while varying the serving-scenario parameters. 
We report the maximum achievable cluster throughput under a TPOT SLO constraint in Figure~\ref{fig:tpot-cl-trend}.

\para{TPOT} Two trends are visible under context length of 512. First, throughput increases as the TPOT constraint becomes more relaxed, because a larger TPOT budget allows the system to use a larger batch size. 
Second, the two clusters achieve nearly identical performance when the TPOT constraint $\leq 20$ ms, but the gap widens quickly as the constraint is relaxed further. At small batch sizes, the $\beta$-term difference between the two clusters is negligible. Once the larger batch sizes are used, the message sizes grow, and the $\beta$-term in the 150~GB/s cluster grows substantially, requiring a much longer TPOT to accommodate larger batches.

\para{Context length} In Figure~\ref{fig:tpot-cl-trend}, the performance gap between the two clusters narrows as context length increases, because the growing KV-cache footprint caps the maximum batch size under fixed memory. For example, with multi-head latent attention (MLA) at context length 8192, the KV cache consumes $\sim$1 GB per request, so a small batch of size=64 nearly fills an H100's 80 GB capacity alongside the model shard.

\subsubsection{Effect of Software Optimizations}
\label{sssec:sw-opt}
We next examine the effect of dual-batch overlap and speculative decoding using the same setup.

\begin{figure}[t]
    \centering

    \begin{subfigure}{0.9\linewidth}
        \centering
        \includegraphics[width=\linewidth]{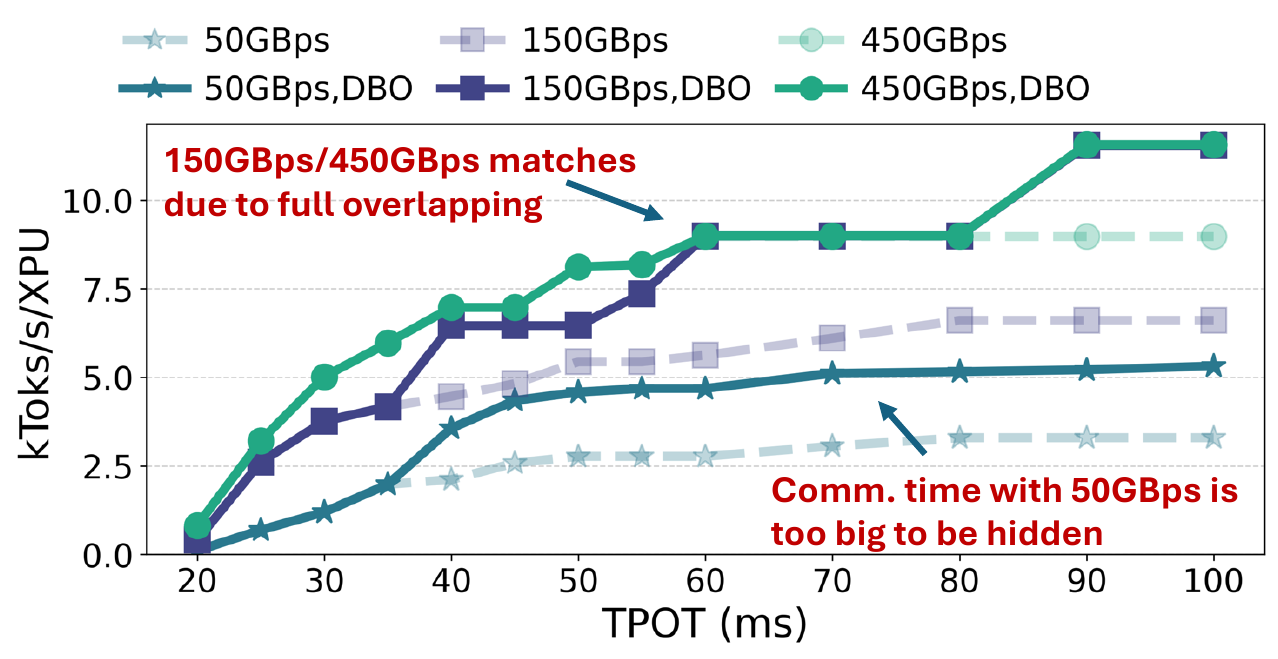}
        \caption{Effect of dual-batch overlap (DBO). No-overlap cases (faded curves) are plotted together for comparison.}
        \label{fig:dbo-effect}
    \end{subfigure}
    \begin{subfigure}{0.85\linewidth}
        \centering
        \includegraphics[width=\linewidth]{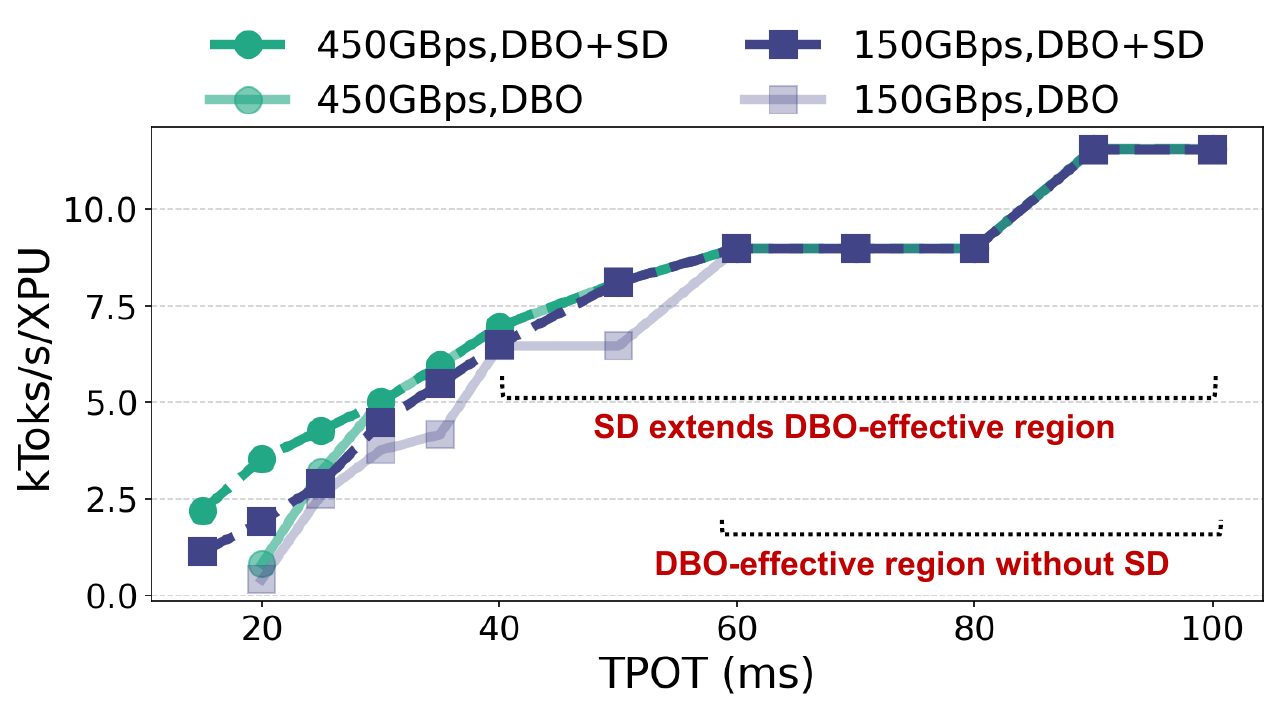}
        \caption{Effect of speculative decoding (SD). DBO-only cases (faded curves) are plotted for comparison.}
        \label{fig:sd-effect}
    \end{subfigure}

    \caption{Effect of software optimizations. The legend represents software optimizations applied in each scenario.}
    \label{fig:optimization-effects}
\end{figure}
\para{DBO} 
Figure~\ref{fig:dbo-effect} shows how DBO changes the throughput-TPOT frontier. The DBO curve shows the best of DBO and the baseline (no overlap) at each TPOT.
As explained in \S\ref{ssec: sw-opt}, when the batch size is small (very low TPOT values), applying DBO instead drops throughput due to low operational intensity, and therefore the DBO curve falls back to the baseline. 
As the TPOT constraint is relaxed, the batch size increases, and DBO begins to improve performance by reducing the exposed communication time. An interesting observation is that the 150 ~GB/s curve with DBO nearly matches the 450 ~GB/s curve once TPOT exceeds 60 ms, indicating that the exposed communication time reaches zero in both clusters. This suggests that 150 ~GB/s is already sufficient to make communication time lower than compute time in this regime.

Does this trend continue as link bandwidth decreases further? The 50 ~GB/s curve shows that higher-bandwidth performance is unreachable even with DBO, because 50 ~GB/s is not sufficient to make compute time exceed communication time.

\para{Spec. Decoding} Figure~\ref{fig:sd-effect} illustrates the effect of adding speculative decoding (SD) on top of DBO. First, SD extends the TPOT range in which DBO is effective. SD increases the effective operating batch size during the verification phase (by $\text{spec}_m$), moving the 40--60~ms region into a regime where OI is large enough for DBO to be effective, thereby narrowing the gap between the 450~GB/s and 150~GB/s systems. In contrast, under DBO alone, a clear gap between the two systems remains in the 40-60~ms region (Figure~\ref{fig:dbo-effect}).

Second, SD improves support for very low-TPOT SLOs, e.g., 15ms. In this setup, the low-TPOT region corresponds to an intermediate batch-size regime: the batch size is not large enough for DBO to be effective, yet it is large enough for the $\beta$-term to become visible. As a result, the 450~GB/s system retains a modest advantage over the 150~GB/s system.

\subsection{Link Bandwidth Comparison} \label{ssec:bw-sweep}
\para{Setup} We compare throughput per cost across scale-up link bandwidths to show the advantage of lower-bandwidth networks under different serving scenarios and software optimizations. Three curves are plotted: Baseline (no optimization), DBO (DBO only), and DBO+SD (both applied).

\begin{figure}[t]
    \centering
    \includegraphics[width=1\linewidth]{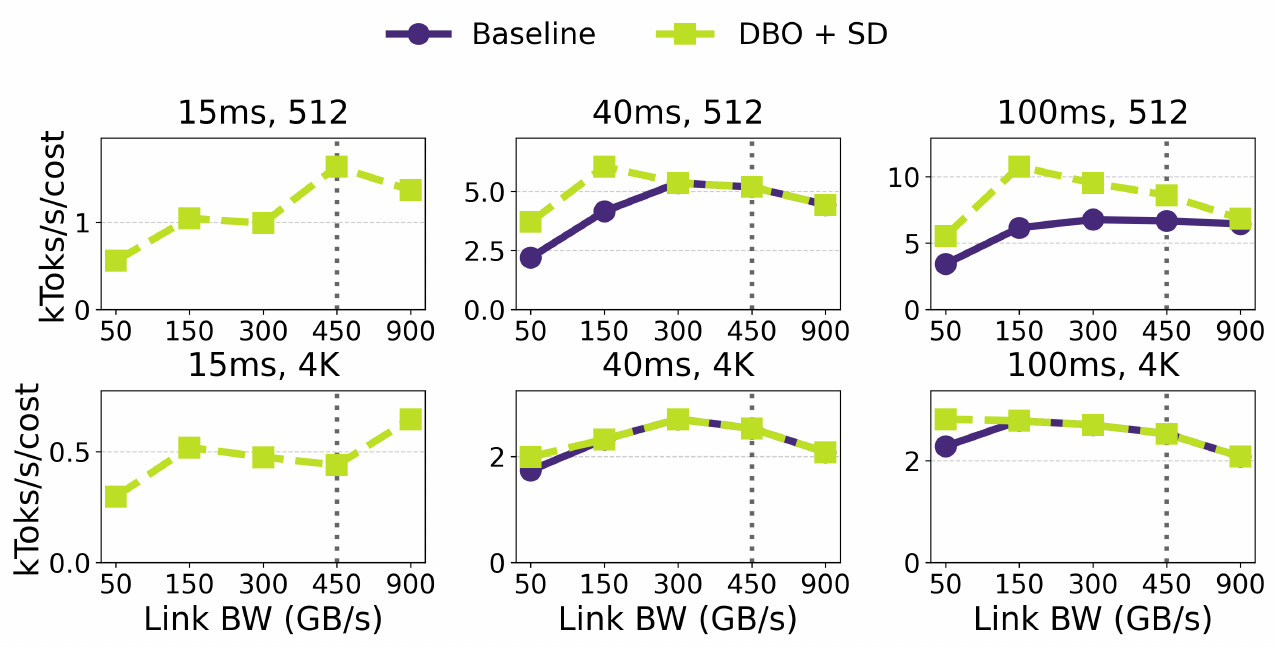}
    \vspace{-0.2in}
    \caption{Throughput per cost versus link BW of scale-up networks in the representative serving scenarios. The vertical line represents the real-world provisioning for Hopper GPUs.}
    \label{fig:link_bw_perf_p_d}
\end{figure}

\para{Result} Figure~\ref{fig:link_bw_perf_p_d} shows the throughput per cost of scale-up networks with different link bandwidths under varying TPOT and context lengths. The first notable trend is that the 1$\times$ provisioning point (450~GB/s) is not the most cost-effective option in most scenarios, even without software optimizations. This suggests that the performance benefit of additional bandwidth becomes marginal beyond a certain point, while the network cost continues to increase due to the higher cost of switches and links that support greater bandwidth. This trend is more pronounced in 4K-context scenarios, as longer context lengths reduce the value of high bandwidth, as discussed in \S\ref{sssec:effect-serving-scenarios}.

\begin{figure}[t]
    \centering
    \includegraphics[width=1\linewidth]{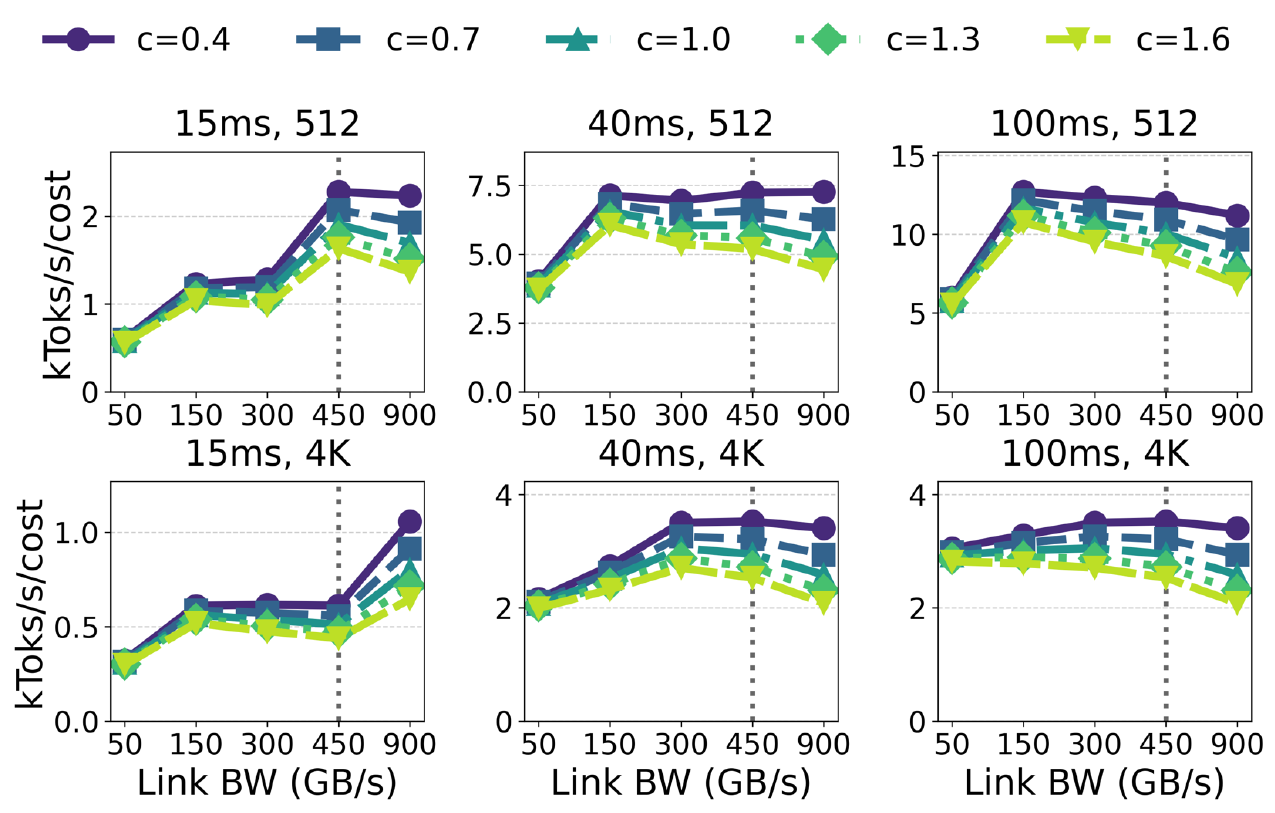}
    \vspace{-0.2in}
    \caption{Throughput per cost versus link BW of scale-up networks, varying cost adjustment factor $c$. All software optimizations are assumed. Lower $c$ represents the case where the network cost is lower than our estimate.}
    \label{fig:link_bw_price_sweep}
\end{figure}

DBO and SD shift the most cost-effective point toward lower bandwidths in the 40–100 ms~TPOT scenarios, with the final sweet spot at 150 or 300 ~GB/s. DBO nearly equalizes high- and low-bandwidth performance in this regime as long as communication time stays below compute time, and SD pushes batch sizes into a DBO-friendly regime in low-TPOT scenarios, letting 150 ~GB/s become more cost-effective than 300 ~GB/s in the 40~ms, 512 scenario. By contrast, 50 ~GB/s is generally not the sweet spot despite its low cost, because its communication time is too large to hide. The exception is the 100 ms, 4K scenario, where per-request compute time is particularly high. Overall, choosing the link-BW sweet spot over the 1$\times$ provisioning baseline improves throughput per cost by 6–27\% across scenarios.

Figure~\ref{fig:link_bw_price_sweep} illustrates how these curves vary with the cost adjustment factor $c$. The sweet spot remains unchanged unless the network cost is significantly lower than our estimate.

\para{\textit{Takeaway:}} \textit{The current link BW provision for scale-up is past the sweet spot for a large fraction of serving workloads when software optimizations are applied.}

\subsection{Topology Comparison} \label{ssec:result-all}
We next compare the cost-effectiveness of the four topologies with XPU network bandwidth fixed.

\subsubsection{Which Topology Is the Most Cost-Effective?} \label{sssec:result-topology}
\begin{figure}[t]
    \centering
    \includegraphics[width=1.0\linewidth]{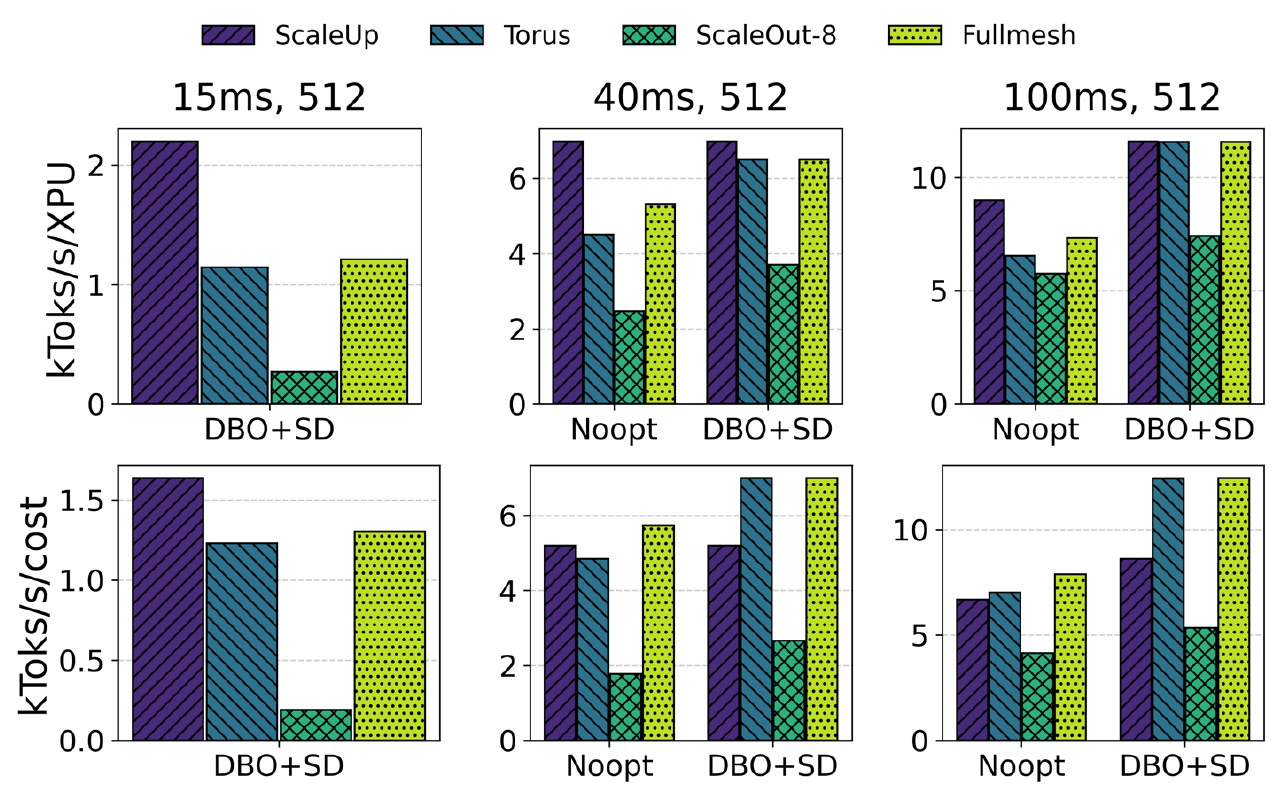}
    \vspace{-0.2in}
    \caption{Throughput per XPU (top) and per unit cost (bottom) on a 64-XPU cluster across topologies. Titles: TPOT, context length. Columns: no optimization (Noopt) and DBO+SD. Noopt at TPOT=15ms misses the SLO.}
    \label{fig:result-all-topologies}
\end{figure}
\begin{figure}[t]
    \centering
    \includegraphics[width=1.0\linewidth]{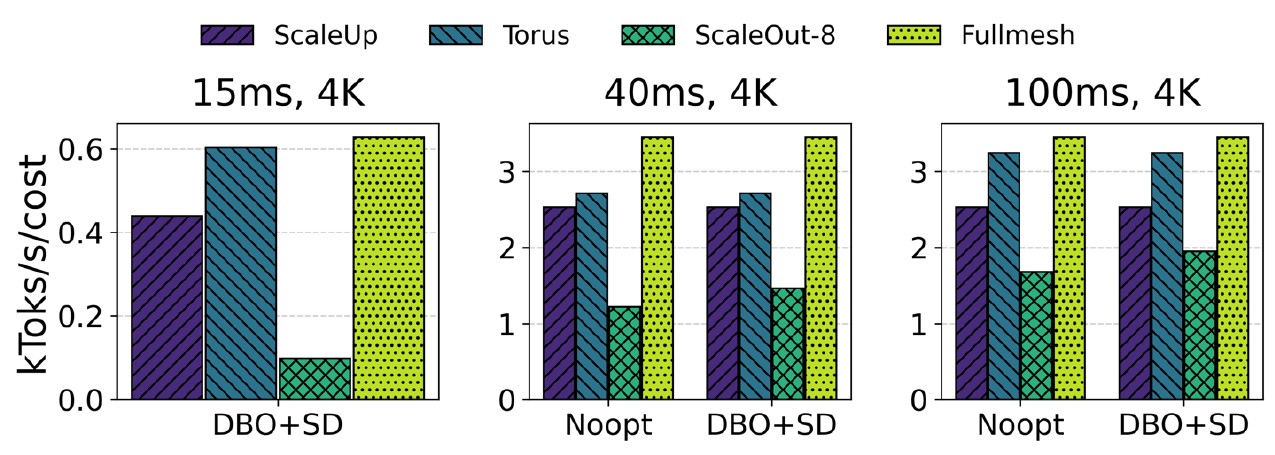}
    \vspace{-0.2in}
    \caption{Throughput per cost in 4K context length scenarios.}
    \label{fig:result-4k-all-topologies}
\end{figure}

\para{Setup} To compare the four topologies under the same provision, we fix the total per-XPU network bandwidth to match the description in \S\ref{ssec: perf-modeling}.
We estimated the throughput per XPU and per unit cost across four topologies of a 64-XPU cluster.
All topologies, switches, and bandwidth parameters follow earlier sections (\S\ref{ssec: background-topologies}, \S\ref{ssec: perf-modeling}, \S\ref{ssec:cost-modeling}).

\para{Result} Figure~\ref{fig:result-all-topologies} shows that switchless torus and full-mesh achieve better throughput per cost than scale-up across all combinations of TPOT SLO and context length. In particular, full-mesh shows the best cost-effectiveness in these serving scenarios.
Under less stringent SLOs (TPOT$\geq$40ms), even the raw performance of low-cost switchless topologies becomes comparable, as DBO hides the disadvantage of switchless topologies, namely, a high $\beta$-term due to indirectly routed traffic and lower per-link bandwidth.

The superiority in throughput per cost in this TPOT regime remains in the long context scenario as shown in Figure~\ref{fig:result-4k-all-topologies}.
As in \S\ref{ssec:bw-sweep}, the raw performance difference between scale-up, torus, and full-mesh decreases in long-context scenarios due to limited batch sizes. The small batch size under long context length leads to small message sizes during allreduce and all-to-all, reducing the $\beta$-term difference between switchless topologies and switch-based topologies. As a result, the topologies with lower costs have higher cost-effectiveness for this workload. DBO is not helpful in this regime because of small batch sizes, even with SD.

In contrast, under strict SLOs (TPOT = 15 ms), which are representative of real-time or reasoning scenarios, scale-up still retains an advantage. The tight TPOT constraint limits the batch size to a range in which DBO is ineffective, making the difference in effective bandwidth apparent. Note that SD is necessary to meet the SLO of TPOT=15ms, and thus, the Noopt (No optimization) column is omitted.

Scale-out is not as effective as other topologies since most MoE LLM traffic crosses the scale-out network, whose bandwidth is 9$\times$ lower than scale-up.

\subsubsection{Is a Larger Cluster More Cost-Effective?} \label{sssec:result-cluster-size}

\begin{figure}[t]
    \centering
    \includegraphics[width=1.0\linewidth]{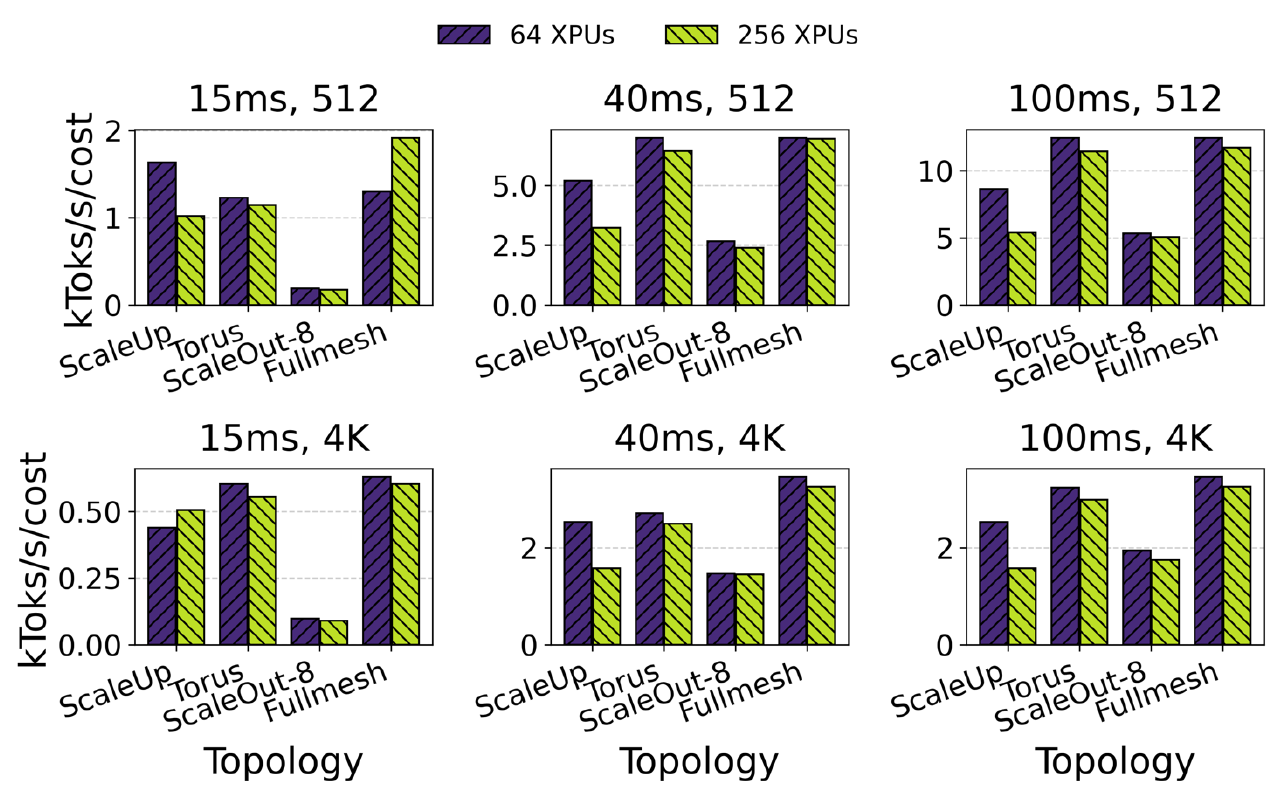}
    \vspace{-0.1in}
    \caption{Token throughput per XPU (top) and per unit cost (bottom). Topologies swept on 64- and 256-XPU clusters.}
    \label{fig:result-cluster-size}
\end{figure}
\begin{figure*}[t]
    \centering
    \includegraphics[width=0.85\textwidth]{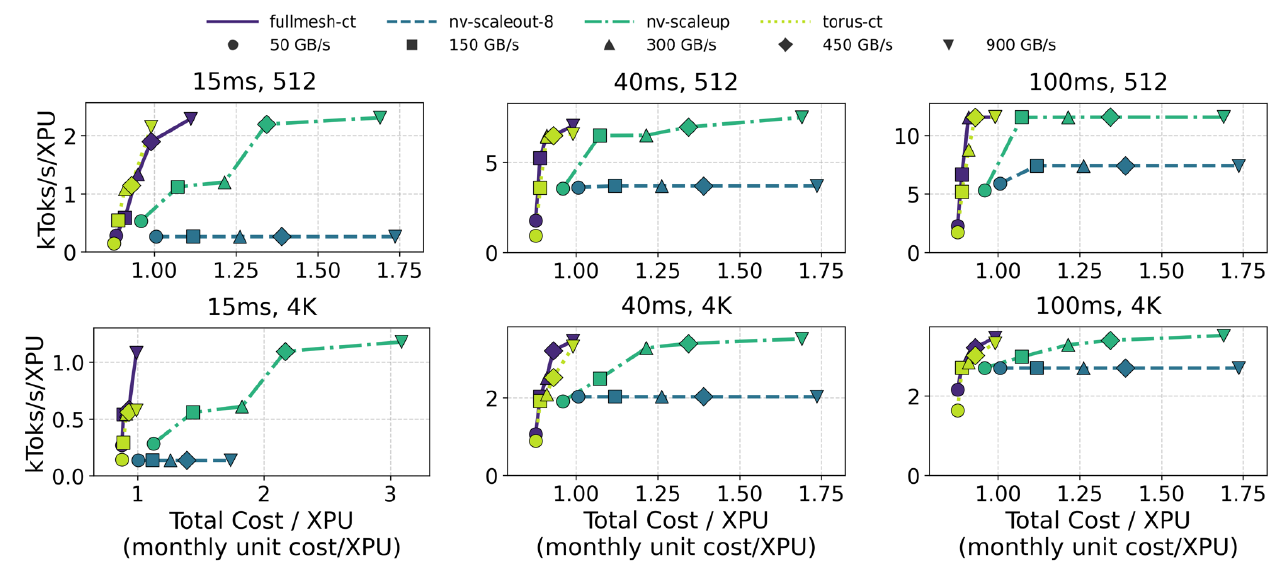}
    \vspace{-0.1in}
    \caption{Performance-cost tradeoff of different XPU networks under varying both topology and link bandwidth. The total cost includes both the monthly XPU and network costs, and the XPU cost is the same across networks. Points that are higher and farther to the left are more cost-effective since the slope from the origin to the point represents cost-effectiveness.}
    \vspace{-0.05in}
    \label{fig:result-best-config}
\end{figure*}

To evaluate the trend at a larger cluster size, we repeat the experiment on 256 XPUs with EP=256, keeping switch radix, per-XPU ports, and bandwidth fixed; only the 3D torus and full-mesh dimensions and full-mesh per-XPU port count increase. Figure~\ref{fig:result-cluster-size} reports results for context length 512.

We first observe that throughput per cost decreases across all topologies in the 40--100ms TPOT regimes. As shown in Table~\ref{table:a2a-alg-coefficients}, enlarging the all-to-all domain increases communication time for every topology. At the same time, the larger cluster size does not provide a meaningful improvement in computation efficiency. As a result, the additional communication overhead directly lowers throughput per cost. This drop is especially pronounced for scale-up. Due to the limited switch radix, scaling past 64 XPUs forces a move from a one-level to a two-level fat-tree, sharply raising switch and link costs and hurting throughput per cost more than other topologies.

Second, some low-TPOT scenarios show substantial improvement in throughput per cost. This gain comes from two factors. First, the slowdown associated with the $\beta$-term does not yet appear in this regime. Second, EP256 can provide better compute efficiency in this regime: moving from EP64 to EP256 reduces the number of experts hosted by each GPU from 4 to 1, thereby reducing MoE execution time at low batch sizes, where computation is bottlenecked by weight loading. Because EP256 does not reduce the total FLOPs per GPU per token, however, it does not improve computation in the large-batch regime, which explains the first observation.

\para{\textit{Takeaway: }} \textit{Switchless topologies are more cost-effective despite being less performant under the 1x bandwidth provision.}

\subsection{Putting It All Together}
\label{ssec:best-config}
Considering all topologies, bandwidths, and cluster sizes, which network offers the best performance per cost? 

\para{Setup} To answer this question, we sweep all parameters and each network on the performance-versus-cost plot in Figure~\ref{fig:result-best-config}. Each point represents (cluster cost per XPU, throughput per XPU), and thus the slope of a hypothetical line that connects the origin to a particular point corresponds to throughput per cost. The lines and shapes correspond to topologies and per-XPU bandwidths. For each topology, the best-performing XPU count and EP are selected per scenario.

\para{Result} Full-mesh forms the Pareto frontier across all serving scenarios, indicating that it is the most cost-effective choice with current-generation XPUs.
Despite the operational difficulties of managing highest number of directly connected links among the evaluated topologies, full-mesh is worth considering for optimizing the cost of clusters when building more clusters is possible.
Torus follows a similar curve but yields much lower throughput at a given bandwidth, while scale-out misses the Pareto frontier entirely.

The main reason the switchless topologies stand out is that all topologies reach throughput saturation, where DBO+SD makes the exposed communication time close to zero, at link bandwidths at or below the current provisioned level (450~GB/s) in most scenarios, echoing the conclusion of the previous section. As a result, the benefit of further investment in more expensive topologies or higher-bandwidth links is only marginal. For example, in the context length=512 and TPOT=$[40, 100]$~ms cases, 1x(450~GB/s), 0.66x (300~GB/s), and 0.33x (150~GB/s) are sufficient to reach the same throughput ceiling for torus, full-mesh, and scale-up, respectively. Notably, the higher-performing the topology is, the lower the bandwidth required to achieve saturation. We use this trend to project how the tradeoff evolves in future generations in \S\ref{ssec:future-projection}.

In low-TPOT, long-context scenarios, switchless topologies either fail to achieve the same ceiling throughput or require more than the current provision to do so, because DBO is ineffective in small-batch regimes. However, they still deliver comparable performance and thus remain more cost-effective because of their lower cost.

Lastly, the raw throughput comparison can be useful when building more clusters is difficult due to real estate or power constraints; the throughput per XPU (y-axis) matters more than the cost efficiency (slope). 
While scale-up achieves the best throughput per XPU across all scenarios, the difference between full-mesh and torus is inconsequential. Scale-up shows the biggest throughput advantage over others in TPOT=15ms, context length=512 scenario. This represents real-time short-context workloads such as auto-completion. The gap between scale-up and others in this scenario can be widened in the future if XPU compute and memory scaling outpaces bandwidth scaling, which we discuss in \S\ref{ssec:future-projection}. Full-mesh can catch up to the performance of scale-up by doubling the link bandwidth while spending less than scale-up. 

\para{\textit{Takeaway: }} \textit{Full-mesh turns out to be Pareto-optimal on performance versus cost for current-generation GPUs.}

\subsection{Will This Trend Continue?} \label{ssec:future-projection}
\begin{table}[h]
\centering
\small
\begin{tabular}{lcc}
\hline
 & Blackwell & Rubin \\
\hline
FP8 FLOPs/s & 2.56$\times$ & 4.49$\times$ \\
Memory Capacity & 2.33$\times$ & 3.60$\times$ \\
Memory Bandwidth & 2.39$\times$ & 6.57$\times$ \\
Per-GPU Link Bandwidth & 2.00$\times$ & 4.00$\times$ \\
\hline
\end{tabular}
\caption{Relative Scaling vs. Hopper (H100 = 1$\times$)}
\vspace{-0.1in}
\label{table:gpu-scaling}
\end{table}

\begin{figure}[t]
    \centering

    \begin{subfigure}{\linewidth}
        \centering
        \begin{overpic}[width=0.9\linewidth]{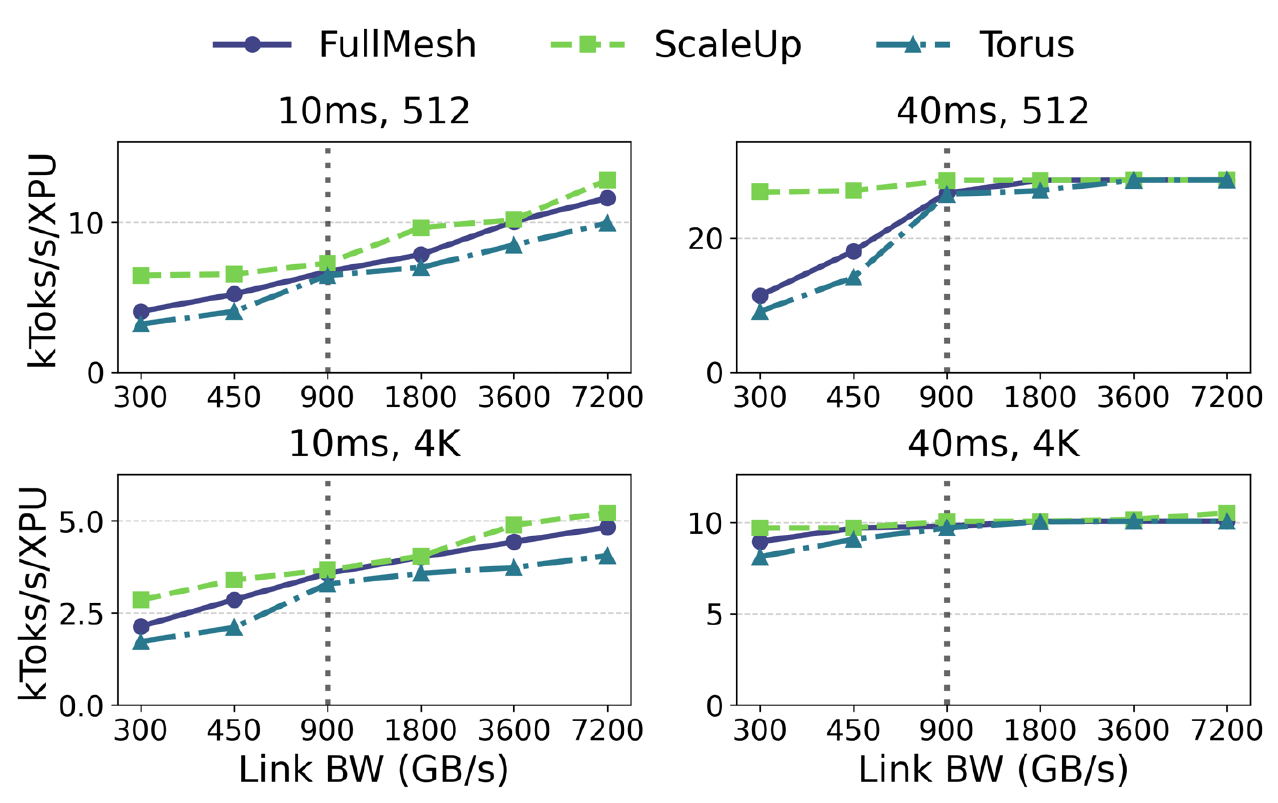}
        \put(-5,30){(a)}
        \end{overpic}
        \phantomcaption
        \label{fig:future-blackwell}
    \end{subfigure}
    \begin{subfigure}{\linewidth}
        \centering
        \begin{overpic}[width=0.9\linewidth]{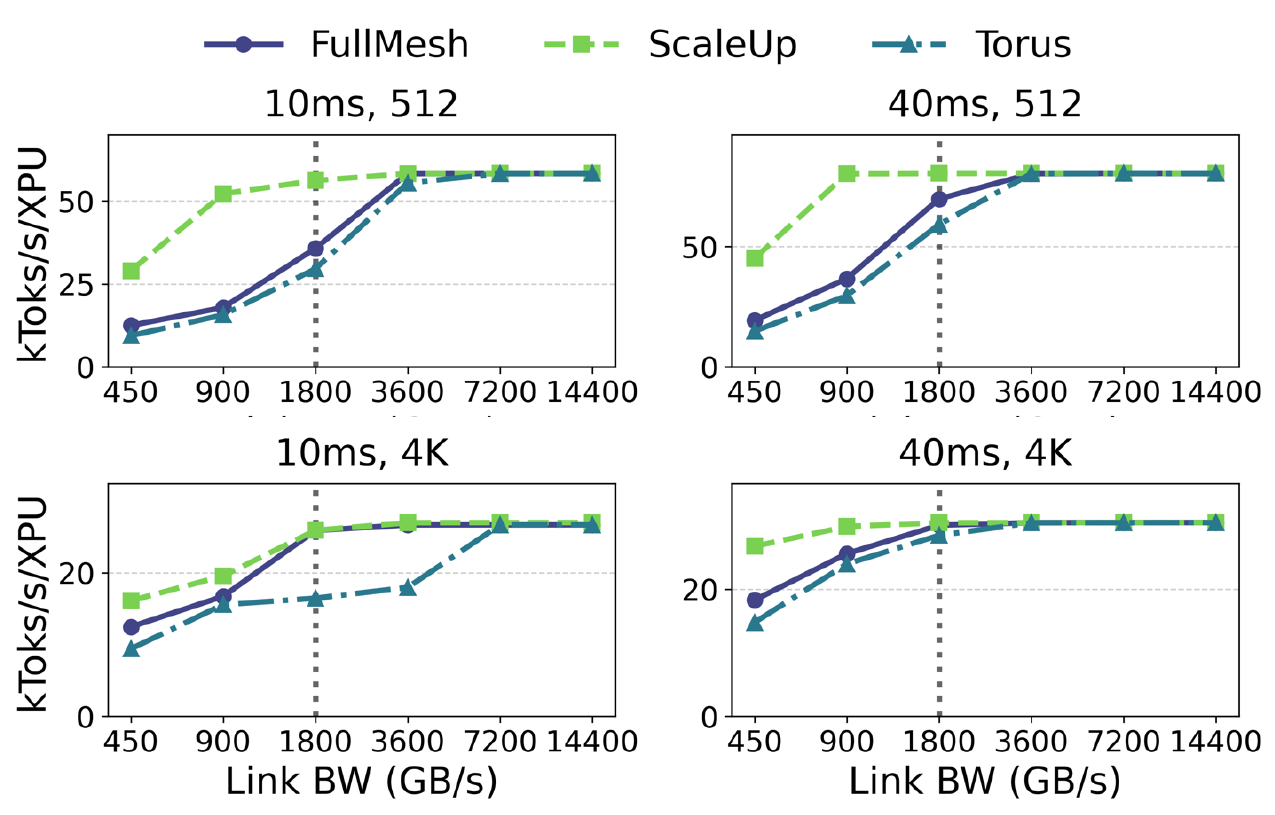}
        \put(-5,30){(b)}
        \end{overpic}
        \phantomcaption
        \label{fig:future-rubin}
    \end{subfigure}
    \caption{Token throughput per GPU in future generations with per-GPU link BW sweep. (a) NVIDIA Blackwell (link BW provision: 900~GB/s)  (b) NVIDIA Rubin (link BW provision: 1800~GB/s).}
    \vspace{-0.2in}
    \label{fig:future}
\end{figure}

Will the cost-effectiveness of switchless topologies continue to hold in future XPU generations? Since cost estimates for future hardware are unavailable, we use the bandwidth required to achieve throughput saturation as a proxy for the cost-effectiveness of different topologies. Our results so far show that, under DBO+SD, a network needs only enough bandwidth to reach the throughput saturation point, beyond which additional bandwidth provides little benefit. Therefore, if the saturating bandwidth of switchless topologies remains below or similar to the bandwidth provision of the next generation, their cost-effectiveness is likely to persist.

The required bandwidth depends on compute time, because saturation occurs when compute time exceeds communication time. To project compute time for future generations, we use a roofline model to estimate the speedup of each kernel relative to H100 based on the projected scaling of FLOPs/s and memory bandwidth, and then apply that speedup to the runtime measured on H100. We use NVIDIA Blackwell and Rubin GPUs as reference points and summarize the assumed scaling trends in Table~\ref{table:gpu-scaling}.

Figure~\ref{fig:future} shows the throughput-bandwidth curves of the three topologies on a 256-XPU system. It presents the results for TPOT targets of 10~ms and 40~ms. We use 10~ms, instead of the 15~ms used in previous sections, since future XPUs are expected to make lower TPOT targets achievable. The 100~ms scenario is omitted as the trend is similar to 40ms.

Figure~\ref{fig:future-blackwell} suggests that the bandwidth provision of Blackwell GPUs (900~GB/s) is sufficient to hold this conclusion. When the context length is long or the SLO is tight, the difference in topology does not lead to a big performance difference owing to small batch sizes. In other scenarios, DBO allows full-mesh and torus to reach the performance of scale-up. In these scenarios, 900~GB/s is sufficient, and the conclusion remains valid even with bandwidth below 900~GB/s.

For Rubin GPUs (1800~GB/s), in Figure~\ref{fig:future-rubin}, the lesson is still true when the context length is long. When the context length is short, scale-up achieves the best throughput and the bandwidth requirement for full-mesh and torus to match scale-up's performance goes up to 3600~GB/s. If the SLO is more relaxed (TPOT$\geq$40ms), full-mesh and torus exhibit competitive performance compared to scale-up as DBO is effective. However, if the SLO is tighter (TPOT=10ms), DBO becomes ineffective even when SD is applied, unless the link bandwidth is higher.

In some Rubin scenarios, the bandwidth requirement for torus and full-mesh to match scale-up increases to 3600~GB/s, owing to the faster scaling of memory (6.57x) relative to link bandwidth (4x) than in previous generations. This suggests that, if this trend continues, the current advantage of the switchless topologies may not persist indefinitely.

\begin{figure}[t]
    \centering
    \includegraphics[width=0.9\linewidth]{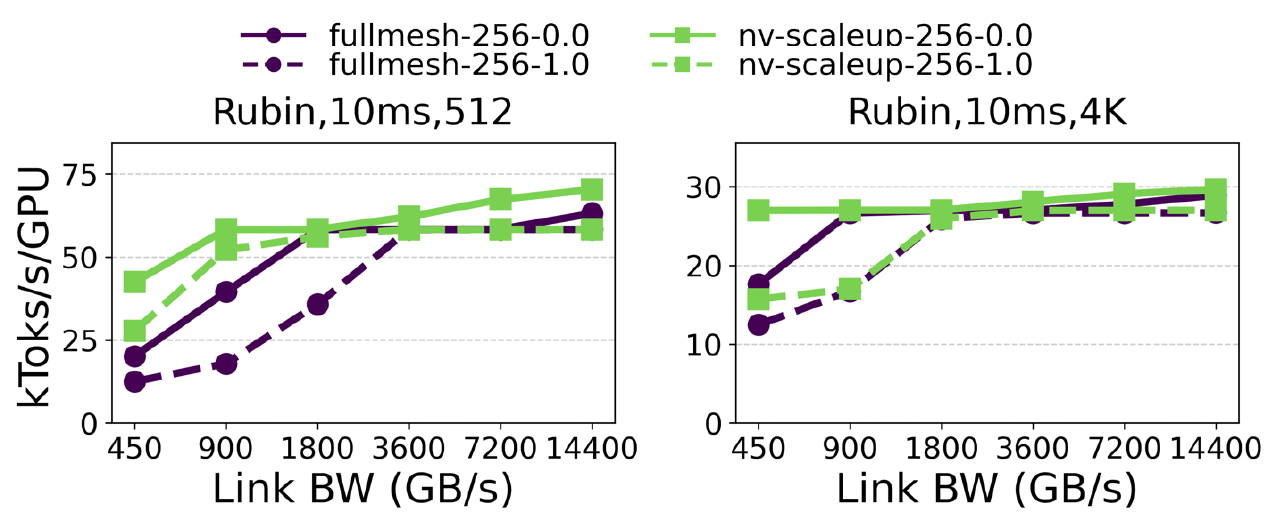}
    \vspace{-0.1in}
    \caption{The effect of the $\alpha$ term in low TPOT regions. 0.0 curves represent the extreme case where $\alpha_{r}$ and $\alpha_{d}$ become zero, while 1.0 curves represent the ones with normal values.}
    \vspace{-0.1in}
    \label{fig:alpha-theta}
\end{figure}

Besides bandwidth scaling, reducing $\alpha_{r}$ and $\alpha_{d}$ in communication time (e.g., via lower software overhead) is helpful when full-mesh is struggling compared to scale-up. Reducing $\alpha_{r}$ and $\alpha_{d}$ allows the $\beta$-term to increase under a certain SLO, leading to a larger message size. As a result, the batch size can reach the region where DBO is effective. The theoretical bound ($\alpha_{r}=\alpha_{d}=0$) of this effect is described in Figure~\ref{fig:alpha-theta} when TPOT=10ms with Rubin GPUs. Under Rubin's bandwidth provision (1800~GB/s), the performance gap between scale-up and full-mesh is removed. In the 4K-context-length case, the bandwidth requirement is even decreased to 900~GB/s. 

\section{Related Work} \label{sec: related-work}

\para{LLM performance modeling} We detail related work in \S\ref{sec: methodology}.

\noindent \textbf{Collective communication algorithms.\hspace{0.5em}}
Many studies explore the topology-agnostic collective algorithms~\cite{gropp,bruck,spread-out,aws-collectives-intra-node,aws-collectives-inter-node}
and topology-specific collective algorithms~\cite{swing,halfring,custom1,custom2,custom3,custom4,supermesh,efficient-a2a-direct-connect}. 
Some works propose topology-aware collective algorithm synthesizers~\cite{taccl,tacos,themis}. Our work builds upon the alpha-beta model of previously suggested algorithms.

\para{XPU network topology.}
We discuss the four commonly used topologies in \S\ref{sec: background}~\cite{nvl72,alibaba-hpn,megascale-bytedance,tpuv4,ub-mesh}. Rail-only~\cite{rail-only} reduces the cost of rail-optimized scale-out Clos networks by removing unnecessary paths in training clusters. This design trades bisection bandwidth for lower cost, making it inadequate for MoE serving, where all-to-all communication dominates. Optical circuit switch (OCS)-based topologies have also been proposed in academia~\cite{topoopt, mixnet}, while we focus only on EPS-based systems.

\para{Network comparison.}
Prior work has studied flow control and collective scheduling across network topologies and bandwidths. General HPC and datacenter studies focus on topology design and analysis~\cite{xpander,jellyfish,tale-of-two,spineless}. ML-specific efforts include flow-level SLO enforcement in serving~\cite{superserve}, GPU-utilization-aware scheduling for training~\cite{crux}, and collective optimization via traffic engineering, load balancing, and auto-tuning~\cite{rethinking-ml-collective,themis,autoccl}. These are purely network-level studies comparing flow completion time; they do not model the compute-communication interaction that determines end-to-end LLM serving performance.

\para{Overlap in LLM serving.}
In addition to DBO, a complementary line of work~\cite{zhang2025janus,megascale-infer} hides all-to-all collectives in LLM serving by pipelining attention and FFN across systems. These approaches share DBO’s core idea: using micro-batching to overlap communication and computation, often more aggressively with three or four micro-batches. We focus on DBO because it is more general and representative of the common techniques used in these systems. Other work overlaps non-all-to-all collectives~\cite{zhu2025nanoflow,wang2022overlap,gond2025tokenweave} in dense LLMs. Since these collectives are not a major bottleneck in MoE serving, we did not consider them.

\section{Conclusion} \label{sec: conclusion}
This paper presents the first systematic, cross-layer comparative analysis of network cost-effectiveness for MoE LLM serving. Low-cost switchless networks, especially 3D full-mesh, consistently deliver 20.6–56.2\% higher performance per cost than scale-up designs across all scenarios studied. This advantage is likely to continue in future XPU generations if interconnect bandwidth keeps pace with compute and memory scaling.

\section*{Acknowledgments}
We thank Jack Snyder and Benjamin Klenk of NVIDIA Research for providing and verifying the NCCL allreduce and all-to-all benchmark data on NVIDIA DGX H100 nodes.
This project is funded by the Microelectronics Commons Program, a DoW initiative.

\bibliographystyle{plain}
\bibliography{references}

\end{document}